\documentclass[11pt,oneside]{article}
\usepackage[utf8]{inputenc}
\usepackage{academicons, lastpage, booktabs}
\usepackage[dvipsnames]{xcolor}
\usepackage[english]{babel}
\usepackage{amsfonts,amssymb,amsmath,amsthm}
\usepackage[absolute,overlay]{textpos}
\usepackage{float}
\usepackage{fancyhdr, graphicx}
\usepackage{enumerate}
\usepackage[margin=1in]{geometry}
\usepackage{tikz}
\usepackage{layout}
\usepackage{titlesec}
\usepackage{titling}
\usepackage{authblk}
\usepackage{xcolor, colortbl}
\usepackage{siunitx} %unités SI
\usepackage{hyperref}
\usepackage{eso-pic}
\usepackage{marginnote}
\usepackage{doi}
\usepackage[skip=6pt plus1pt, indent=20pt]{parskip}
\usepackage{gensymb}
\usepackage{multirow}

\usepackage[version=4]{mhchem}

\definecolor{coolblack}{rgb}{0.0, 0.18, 0.30}
\definecolor{carmine}{rgb}{0.59, 0.0, 0.09}
\definecolor{myblue}{rgb}{0.0, 0.48, 0.65}
\definecolor{cadetgrey}{rgb}{0.57, 0.64, 0.69}
\definecolor{lightgray}{gray}{0.9}

\hypersetup{
    colorlinks=true,
    linkcolor=coolblack,
   	filecolor=magenta,      
    urlcolor=cyan,
    citecolor=coolblack,
    pdfhighlight=/P,
                 }
                                  
\usepackage{libertinus}
\usepackage[T1]{fontenc}
\usepackage{orcidlink} %for the orcid logo print and link

\usepackage{caption}
\usepackage{fancyhdr}

\titleformat*{\section}{\Large\bfseries\color{black}}
\titleformat*{\subsection}{\large\bfseries\color{black}}
\titleformat*{\subsubsection}{\bfseries\color{black}}

\makeatletter
\renewcommand\@makefntext[1]{\leftskip=0em\hskip0em\@makefnmark#1}
\makeatother
\newcommand\blfootnote[1]{%
  \begingroup
  \renewcommand\thefootnote{}\footnote{#1}%
  \addtocounter{footnote}{-1}%
  \endgroup
}

\providecommand{\keywords}[1]
{
  \small	
  \textbf{\textsf{Keywords---}} #1
}

\renewcommand{\tableautorefname}

\newcommand{\mycomment}[1]{}
\newcommand*\mytitle{Dynamics of an impulse dielectric barrier discharge in pure ammonia gas using electrical characteristics and imaging analysis}
\newcommand*\myauthor{Jean-Marie-Desiree \textit{et al.}}
\newcommand*\mycorrmail{stephane.cuynet@univ-lorraine.fr} %à compléter
\newcommand*\myDOI{10.46298/ops.17129}
\newcommand*\myarxiv{2512.09454} 
\newcommand*\myHAL{-}
\newcommand*\myzenodo{-} 
\newcommand*\myvolume{2} 
\newcommand*\myyear{2026} 
\newcommand*\mynumber{4 } 
\newcommand*\mydaterec{December 16, 2025} 
\newcommand*\mydaterevised{April 15, 2026}
\newcommand*\mydateaccepted{July 13, 2026} 
\newcommand*\mydatepublished{October 1, 2026}

\newlength{\myshift}
\usepackage[firstpage=true]{background}
\backgroundsetup{contents={\includegraphics[width=1.\paperwidth]{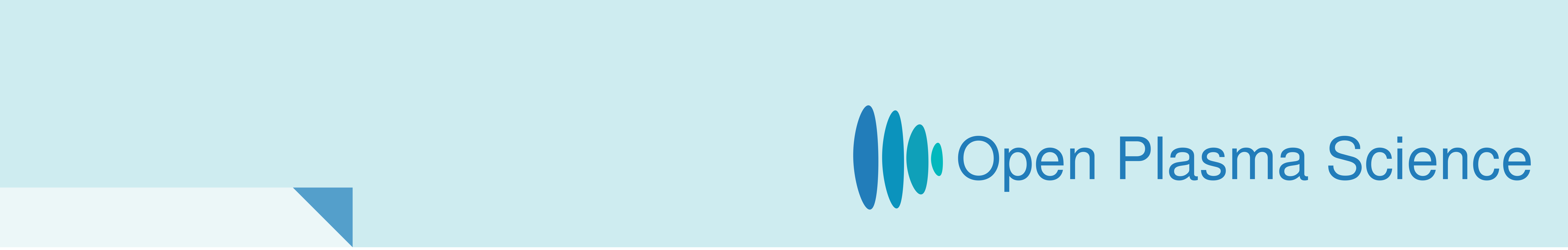}},scale=1,placement=top,opacity=1}

\pretitle{\vspace{-00pt}      \begin{flushleft}   \Large  \color{black} \selectfont  } % Horizontal rule before the title
     
\title{\textbf{\mytitle}}

\posttitle{\par \end{flushleft} } % Whitespace under the title

\preauthor{\vspace{-00pt} \begin{flushleft}  \color{black} \selectfont} % Horizontal rule before the title

\author[1]{Ronny Jean-Marie-Désirée\orcidlink{0000-0002-1882-9138}} 
\author[2]{Aymane Najah\orcidlink{0000-0002-7545-8388}}
\author[1]{Ludovic De Poucques\orcidlink{0000-0001-5707-5490 }} 
\author[1]{Stéphane Cuynet\orcidlink{0000-0002-0726-6093}}

\affil[1]{\small Institut Jean Lamour, Université de Nancy, Nancy, France}
\affil[2]{\small Groupe de Recherches sur l’Énergétique des Milieux Ionisés, Université d’Orléans, Orléans, France}

\postauthor{\par\end{flushleft}} % Whitespace under the title

\date{}

\begin{document}

\maketitle
\thispagestyle{empty}
\blfootnote{$^*$ Corresponding author: \textsf{\href{\mycorrmail}{\mycorrmail}}}
%\blfootnote{$^{\dag}$ Now at: elsewhere} %à compléter/supprimer si non nécessaire
\blfootnote{Cite as: \myauthor, \mytitle, \textit{Open Plasma Science}  \myvolume , \mynumber (\myyear), doi: \myDOI}

%%%%%%%%%%%%%%%%%%%%%%%%%%%%%%%%%%%%%%%%%%%%%%%%%%%%%%%%%%%%%%%%%%%%%%%%%%%%%%%%%%%%%%%%%%
%LATERAL PANNEL OF THE FIRST TITLE PAGE AUTOMATIC JUST ADJUST \mywidth above
%%%%%%%%%%%%%%%%%%%%%%%%%%%%%%%%%%%%%%%%%%%%%%%%%%%%%%%%%%%%%%%%%%%%%%%%%%%%%%%%%%%%%%%%%%

\reversemarginpar

\marginnote{\begin{flushleft}
\sf \bf \footnotesize \color{coolblack} History
\end{flushleft}}[-\myshift]
\addtolength{\myshift}{-0.5cm} 
\marginnote{
\begin{flushleft} \footnotesize
Received \mydaterec
\end{flushleft}}[-\myshift]

\addtolength{\myshift}{-0.5cm} 
\marginnote{
\begin{flushleft} \footnotesize
Revised \mydaterevised
\end{flushleft}}[-\myshift]

\addtolength{\myshift}{-0.5cm} 
\marginnote{
\begin{flushleft} \footnotesize
Accepted \mydateaccepted
\end{flushleft}}[-\myshift]

\addtolength{\myshift}{-0.5cm} 
\marginnote{
\begin{flushleft} \footnotesize
Published \mydatepublished
\end{flushleft}}[-\myshift]

\addtolength{\myshift}{-1cm} 

\marginnote{\begin{flushleft}
\sf \bf \footnotesize \color{coolblack} Identifiers
\end{flushleft}}[-\myshift]
\addtolength{\myshift}{-0.5cm} 
\marginnote{
\begin{flushleft} \footnotesize
DOI \href{https://doi.org/\myDOI}{\myDOI} %à compléter/effacer si besoin
\end{flushleft}}[-\myshift]
\addtolength{\myshift}{-0.5cm} 
\marginnote{
\begin{flushleft} \footnotesize
HAL \href{https://hal.science/\myHAL}{\myHAL} %à compléter/effacer si besoin
\end{flushleft}}[-\myshift]
\addtolength{\myshift}{-0.5cm} 
\marginnote{
\begin{flushleft} \footnotesize
ArXiv \href{https://arxiv.org/abs/\myarxiv}{\myarxiv} %à compléter/effacer si besoin
\end{flushleft}}[-\myshift]
\addtolength{\myshift}{-0.5cm} 
\marginnote{
\begin{flushleft} \footnotesize
Zenodo \href{https://https://zenodo.org/records/\myzenodo}{\myzenodo} %à compléter/effacer si besoin
\end{flushleft}}[-\myshift]

\addtolength{\myshift}{-1cm} 
\marginnote{\begin{flushleft}
\sf \bf \footnotesize \color{coolblack} Supplementary Material
\end{flushleft}}[-\myshift]
\addtolength{\myshift}{-0.5cm} 
\marginnote{
\begin{flushleft} \footnotesize
- %\href{https://arxiv.org}{Something Linky} %à compléter/effacer si besoin
\end{flushleft}}[-\myshift]

\addtolength{\myshift}{-1cm} 
\marginnote{\begin{flushleft}
	\sf \bf \footnotesize \color{coolblack} Licence
\end{flushleft}}[-\myshift]
\addtolength{\myshift}{-0.5cm} 
\marginnote{
\begin{flushleft} \footnotesize
	\href{https://creativecommons.org/licenses/by/4.0/}{CC BY 4.0} 
\end{flushleft}}[-\myshift]
\addtolength{\myshift}{-0.5cm}
\marginnote{
\begin{flushleft} 
	\includegraphics[width=0.55\marginparwidth]{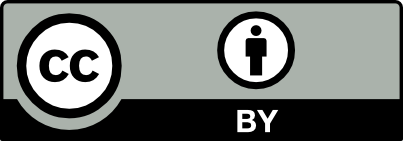}
\end{flushleft}}[-\myshift]
\addtolength{\myshift}{-1.25cm} 
\marginnote{
	\begin{flushleft} \footnotesize
		\copyright  The Authors
\end{flushleft}}[-\myshift]

\addtolength{\myshift}{11.75cm} %NE PAS TOUCHER ICI, TOUCHER + HAUT
\marginnote{\begin{flushleft}
\sf \footnotesize \color{coolblack} \textsc{\textbf{Vol. \myvolume, No. \mynumber (\myyear)}}
\end{flushleft}}[-\myshift]

%%%%%%%%%%%%%%%%%%%%%%%%%%%%%%%%%%%%%%%%%%%%%%%%%%%%%%%%%%%%%%%%%%%%%%%%%%%%%%%%%%%%%%%%%%
%%%%%%%%%%%%%%%%%%%%%%%%%%%%%%%%%%%%%%%%%%%%%%%%%%%%%%%%%%%%%%%%%%%%%%%%%%%%%%%%%%%%%%%%%%
%FILL IT BELOW THE TYPE OF COMMUNICATION : REGULAR ARTICLE / LETTER / SPECIAL ISSUE / TOPICAL REVIEW
%%%%%%%%%%%%%%%%%%%%%%%%%%%%%%%%%%%%%%%%%%%%%%%%%%%%%%%%%%%%%%%%%%%%%%%%%%%%%%%%%%%%%%%%%%
%%%%%%%%%%%%%%%%%%%%%%%%%%%%%%%%%%%%%%%%%%%%%%%%%%%%%%%%%%%%%%%%%%%%%%%%%%%%%%%%%%%%%%%%%%

\addtolength{\myshift}{-1cm} %<---- supprimer si besoin la parenthèse chiffrée
  
\marginnote{\begin{flushleft}
\sf \normalsize \color{coolblack} \textsc{\textbf{Regular article}}
\end{flushleft}}[-\myshift]

%%%%%%%%%%%%%%%%%%%%%%%%%%%%%%%%%%%%%%%%%%%%%%%%%%%%%%%%%%%%%%%%%%%%%%%%%%%%%%%%%%%%%%%%%%
%ABSTRACT
%%%%%%%%%%%%%%%%%%%%%%%%%%%%%%%%%%%%%%%%%%%%%%%%%%%%%%%%%%%%%%%%%%%%%%%%%%%%%%%%%%%%%%%%%%
\begin{flushleft}
\textbf{\textsf{Abstract}}
\vspace{5pt} 
\end{flushleft}
%%%%%%%%%%%%%%%%%%%%%%%%%%%%%%%%%%%%%%%%%%%%%%%%%%%%%%%%%%%%%%%%%%%%%%%%%%
%TEXT of the abstract below
%%%%%%%%%%%%%%%%%%%%%%%%%%%%%%%%%%%%%%%%%%%%%%%%%%%%%%%%%%%%%%%%%%%%%%%%%%

A glow nanosecond discharge from a plane-to-plane impulse dielectric barrier discharge (iDBD) with ammonia gas has been characterized by employing fast imaging and electrical diagnostics. More precisely, the aim of this study is to investigate the dynamics of the discharge establishment under various conditions of applied voltage, pressure, and gas gap. The comparison between the current measurements and the image analysis reveals a strong correlation between the fast excitation and ionisation wave velocity and the current rise rate. This correlation has been found only for diffuse mode discharge since a front wave could be clearly defined, denoted as the luminous propagation front (LPF). Furthermore, this correlation is supported by a proportionality factor of $1.5 \times 10^{-3}$ which is systematic over the studied conditions. Further investigations are considered to evaluate the relevance of such a value over more parameters.

%%%%%%%%%%%%%%%%%%%%%%%%%%%%%%%%%%%%%%%%%%%%%%%%%%%%%%%%%%%%%%%%%%%%%%%%%%
%END ABSTRACT
%space below b4 the keywords
%%%%%%%%%%%%%%%%%%%%%%%%%%%%%%%%%%%%%%%%%%%%%%%%%%%%%%%%%%%%%%%%%%%%%%%%%%

\vspace{20pt} 

%%%%%%%%%%%%%%%%%%%%%%%%%%%%%%%%%%%%%%%%%%%%%%%%%%%%%%%%%%%%%%%%%%%%%%%%%%%%%%%%%%%%%%%%%%
%keywords
%%%%%%%%%%%%%%%%%%%%%%%%%%%%%%%%%%%%%%%%%%%%%%%%%%%%%%%%%%%%%%%%%%%%%%%%%%%%%%%%%%%%%%%%%%

\begin{flushleft}

\keywords{Impulse Dielectric Barrier Discharge, Fast Ionisation Waves, fast imaging measurements}

\end{flushleft}

\newpage
%%%%%%%%%%%%%%%%%%%%%%%%%%%%%%%%%%%%%%%%%%%%%%%%%%%%%%%%%%%%%%%%%%%%%%%%%%%%%%%%%%%%%%%%%%
%FOOT NOTE AND HEADER DOCUMENT
%%%%%%%%%%%%%%%%%%%%%%%%%%%%%%%%%%%%%%%%%%%%%%%%%%%%%%%%%%%%%%%%%%%%%%%%%%%%%%%%%%%%%%%%%%

\newgeometry{ left=20mm, right=20mm,  bottom=3.5cm }

\pagestyle{fancy}
\setlength{\headheight}{15pt}
\fancyhead{} % clear all header fields
\fancyhead[L]{\footnotesize{\href{https://doi.org/\myDOI}{\myDOI}}} %à compléter/modifier ?
\fancyhead[R]{\footnotesize{\myauthor}}

\fancyfoot{} % clear all footer fields
\fancyfoot[R]{\color{coolblack}{ \thepage \hspace{1pt} | \pageref{LastPage}}}
\fancyfoot[C]{\footnotesize{\color{black} Open Plasma Science \textbf{\myvolume}, No.\mynumber (\myyear)}}

\renewcommand{\headrulewidth}{0.4pt}
\renewcommand{\footrulewidth}{0.4pt}
\renewcommand{\footruleskip}{5pt}
%\addtolength{\headwidth}{\marginparwidth}
\renewcommand\footrule{\hrule width\textwidth}
\renewcommand\headrule{\hrule width\textwidth}

%%%%%%%%%%%%%%%%%%%%%%%%%%%%%%%%%%%%%%%%%%%%%%%%%%%%%%%%%%%%%%%%%%%%%%%%%%%%%%%%%%%%%%%%%%
%TABLE OF CONTENTS
%%%%%%%%%%%%%%%%%%%%%%%%%%%%%%%%%%%%%%%%%%%%%%%%%%%%%%%%%%%%%%%%%%%%%%%%%%%%%%%%%%%%%%%%%%

\tableofcontents

%%%%%%%%%%%%%%%%%%%%%%%%%%%%%%%%%%%%%%%%%%%%%%%%%%%%%%%%%%%%%%%%%%%%%%%%%%%%%%%%%%%%%%%%%%
%DOCUMENT
%%%%%%%%%%%%%%%%%%%%%%%%%%%%%%%%%%%%%%%%%%%%%%%%%%%%%%%%%%%%%%%%%%%%%%%%%%%%%%%%%%%%%%%%%%

\section{Introduction}

The observation, characterization and simulation of excitation and ionisation wave propagation have sparked significant interest in the scientific community for several decades. This propagation phenomenon has been studied in various plasma configurations, such as atmospheric pressure plasma jets and dielectric-barrier discharges (DBD) working at relatively high pressure (more than ten thousand pascal).

In the case of volume dielectric-barrier discharge arrangements \cite{brandenburg_dielectric_2017, gibalov_dynamics_2012}, the study of these discharges exposes various operating modes based on the experimental conditions. These modes can be classified into three general categories \cite{kogelschatz_filamentary_2002, massines_comparison_1998, hoft_exploring_2023, belinger_transition_2017, cross_pattern_1993, ouyang_nonlinear_2018}: filamentary, diffuse (also referred to as homogeneous) and patterned (also called self-organised). From a phenomenological perspective, the filamentary discharge mode is characterized by streamer breakdown microdischarges, randomly distributed in time and space over a dielectric plate. On the contrary, the diffuse discharge mode refers to discharges that extend over the entire electrode surface generally arising from a Townsend breakdown according to the literature. More precisely, diffuse modes are described in the literature either as Townsend-type discharges (APTD) or glow-type discharges (APGD). The Townsend-type discharge is characterized by a localised emission region near the anode and an exponential increase in electron density toward the anode \cite{adler_formation_2000, brandenburg_diffuse_2005, muller_various_1996}. By contrast, the glow-type discharge is characterized by a light-emission distribution that resembles that of a low-pressure glow discharge \cite{navratil_comparative_2005, mangolini_radial_2002, obradovic_spectroscopic_2008, ivkovic_electric_2012}. A common interpretation of such a structure is a relatively slow effective ionisation frequency due to multi-step ionisation processes, such as Penning ionization, which prevent arc formation \cite{boeuf_plasma_2003, massines_physics_2003}. The patterned discharge mode, characterized by stable, localised, and reproducible discharges both spatially and temporally, represents a variant where radial diffusion is somehow inhibited or at least differ from the diffuse mode \cite{purwins2011, qiao_mechanism_2016, zhang_multimethods_2019, stollenwerk2006}. 

However, the given descriptions of these different modes essentially deal with a well-established discharge, meaning that a luminous channel has formed across the entire inter-electrode space. Therefore, the understanding of the transient physical processes occurring during the discharge establishment is crucial to complete this description. Furthermore, a deeper insight into these transient mechanisms could enable better control of the desired modes by adjusting experimental conditions such as applied voltage characteristics, pressure, and inter-electrode distance. Indeed, various articles highlight local electron mechanisms, which result in an electric field confinement effect \cite{ouyang_nonlinear_2018, stollenwerk2006} named electron focusing effect. This effect is induced by the presence of surface and volume charges originating from the previous discharge and which self-organised during the very first discharges. As an example, J. Ouyang and coworkers \cite{ouyang_nonlinear_2018, qiao_mechanism_2016, zhang_multimethods_2019, li_comparing_2016} experimentally illustrated the influence of discharge periodicity with a plane-to-plane DBD configuration, thereby demonstrating the existence of transitions from a diffuse mode to a self-organised discharge mode depending on the working frequency.

The aim of the present work is to characterize the luminous propagation front (LPF) generated by an impulse dielectric barrier discharge (iDBD) in an ammonia gas using an iCCD camera and to compare its characteristics with electrical measurements. The purpose of such a study is to highlight a relationship between the temporal voltage-current evolutions and those of the LPF, through the electron flux propagation. Besides, this iDBD cell has already been used in order to investigate the axial electric field employing E-FISH laser technique coupled with voltage and current measurements \cite{Jean-Marie-Desiree_2024}. Similarly to our previous study \cite{Jean-Marie-Desiree_2024}, the LPF characteristics have been studied over three sets of the following experimental parameters: the applied voltage, the working gas pressure, and the gas gap in-between dielectrics of the iDBD cell.

\section{Experimental setup}\label{experimental_setup}

The experimental setup section briefly describes the iDBD cell, the power supply used to generate the discharge, and the iCCD camera employed to characterize the latter. A complete description of the discharge reactor and the power supply is reported in our previous work \cite{Jean-Marie-Desiree_2024}.

\begin{figure}[h!]
	\centering
	\includegraphics[width=0.6\linewidth]{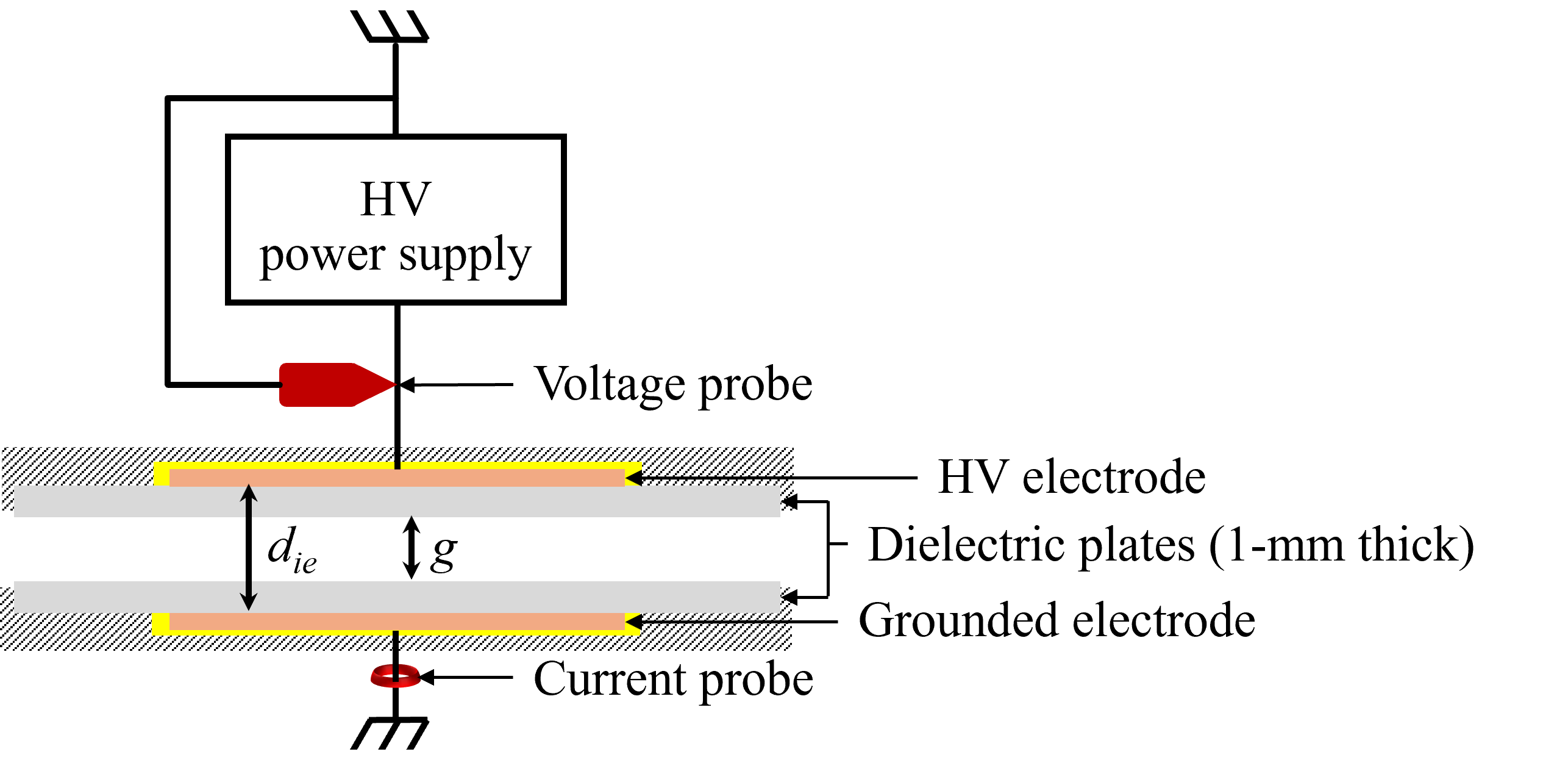}
	\caption{Experimental schematic of the iDBD cell.}
	\label{fig1}
\end{figure}

The iDBD cell is designed in a planar configuration, featuring two solid-state dielectrics in the inter-electrode gap made of aluminium nitride and separated by pure ammonia gas, as schematically illustrated in figure \ref{fig1}. One dielectric is attached to the upper copper electrode which is connected to the high-voltage supply while the second dielectric is attached to the lower copper electrode which is grounded. Thus, the inter-electrode distance $d_{ie}$ corresponds to the sum of the two 1-mm thick dielectric layers and the gas gap denoted as \emph{g}: $d_{ie} = 2 + g$ [mm]. The discharge itself is driven by a homemade bipolar high-voltage power supply at 4 kHz with a symmetric square pulse voltage. Voltage and current measurements were performed around the iDBD cell, with a voltage probe (CalTest Electronics CT4028, bandwidth up to 220 MHz) positioned on the high-voltage copper electrode and a current probe (MagneLab CT1.0, bandwidth up to 500 MHz) on the grounded copper electrode. 

To study the light emission dynamic from the discharge, a picosecond high-speed gated iCCD camera from Stanford Computer Optics was positioned facing the iDBD cell. This camera is equipped with an EF 100 mm f/2.8 Canon macro lens giving a depth of field of approximately 5 mm, which is significantly smaller than the dimensions of the square electrodes ($\sim 3.2 \times 3.2~$cm$^2$). Moreover, the iCCD camera is synchronised with the applied voltage and gated at 1 ns. The resulting image is obtained by accumulating 5000 broadband emission shots from the discharge, with a temporal resolution of 1 ns. These images are then phased with the electrical characteristics by associating the temporal evolution of the discharge current $I_{dis}$ with that of the integrated intensity over the gas gap, as both exhibit a strong correlation.

\begin{table}[h!]
\centering
\caption{Overview of the main parameters among the studied conditions}
\label{table1}
\begin{tabular}{ |c|c|c|c| }
	\cline{2-4} %remplace \hline
	\multicolumn{1}{c|}{} & \multicolumn{3}{|c|}{Influence of} \\%cellules fusionnées au niveau de la ligne
	\cline{2-4}
	\multicolumn{1}{c|}{} & Voltage (sect. \ref{voltage}) & Pressure (sect. \ref{pressure}) & Gas gap (sect. \ref{gas_gap}) \\
	\hline
	$V$ (kV$_{pp}$) & \textcolor{red}{4 -- 8} & 6 & 6 \\
	\hline
	$p~(\times 10^4 Pa)$ & 1 & \textcolor{red}{0.8 -- 2.0} & 1 \\
	\hline
	\emph{g} (mm) & 3 & 3 & \textcolor{red}{3 -- 9} \\
	\hline
\end{tabular}
\end{table}

Electrical measurements and iCCD ultra-fast camera have been carried out for both negative transition NT (from a higher to a lower voltage) and positive transition PT (from a lower to a higher voltage), for each condition listed in table \ref{table1}. One can note that only the measurements obtained with the negative transition is presented and discussed in this paper. Since the applied voltage is symmetrical, the behaviour described with NT is similar and symmetrical with PT.

\section{Results and discussion}
The results and discussion section begins with a general description of the typical observations made in this discharge configuration in ammonia gas, with a reference conditions: 8 kV$_{pp}$, 1 $\times$ 10$^4$ Pa and 3-mm gap. This first approach is followed by a more detailed study that deals with the correlation between imaging analysis and electrical analysis for the different studied operating parameters (cf. table \ref{table1}).

\subsection{General observations and descriptions}\label{gen_obvs}
The goal of this subsection is to present general observations and clarify the type of regime that characterizes the pulsed discharges implemented here.

\begin{figure}[h!]
	\centering
	\includegraphics[width=0.7\linewidth]{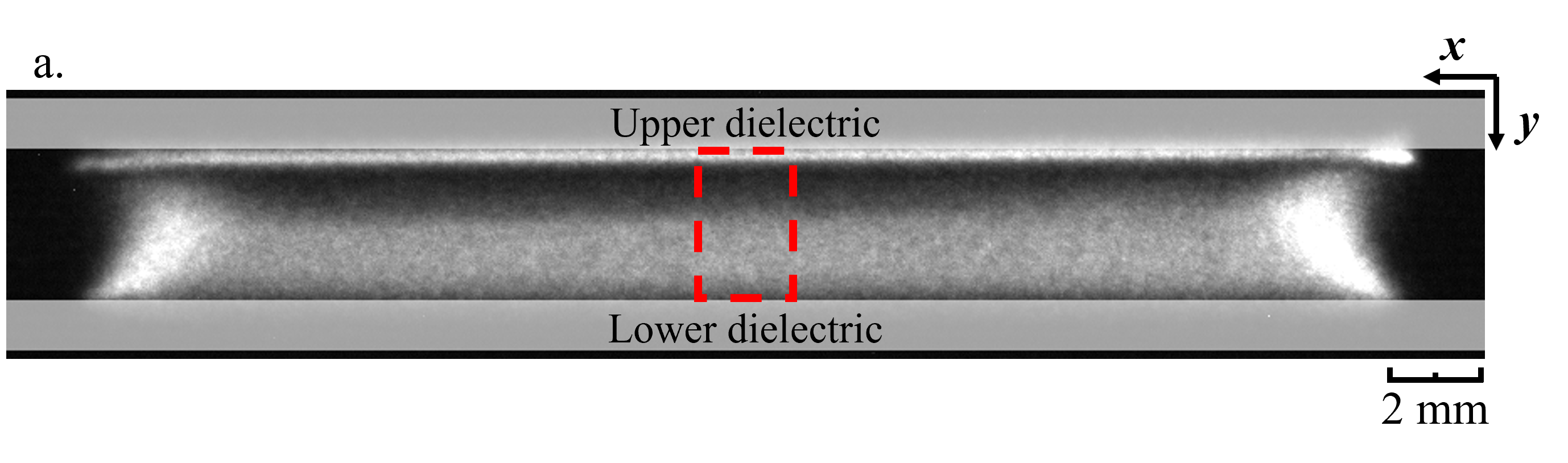}
	\hfill
	\includegraphics[width=0.7\linewidth]{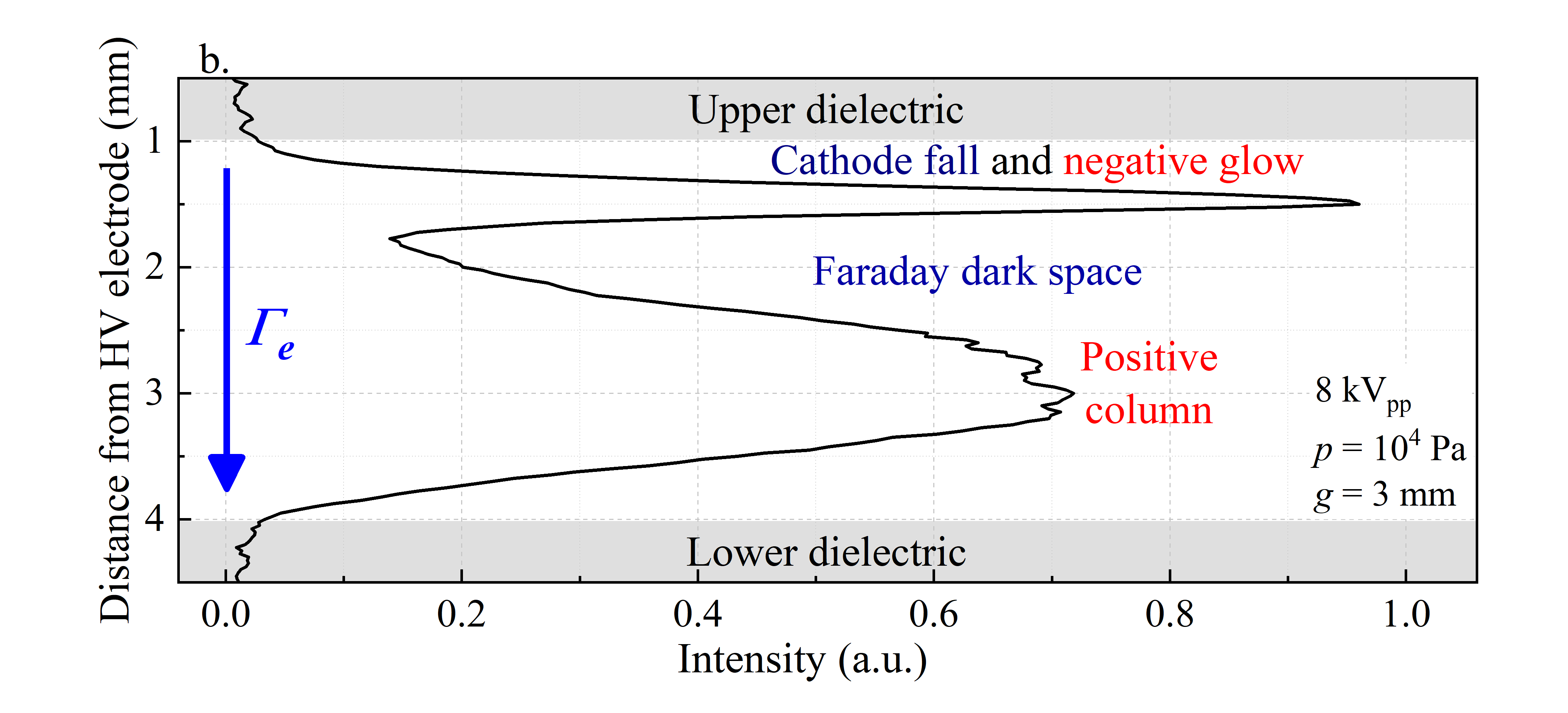}
	\caption{(a) iCCD ultra-fast camera image of iDBD discharge in pure ammonia at 80 ns from the beginning of the NT ($t_0$).\\ (b) Distribution of the normalised light intensity along the \emph{y}-axis, integrated over 2 mm around the centre, as indicated by the red rectangle in (a). Measurements were taken at $10^4$ Pa, 8 kV$_{pp}$, and \emph{g} = 3 mm.}
	\label{fig2}
\end{figure}

Figure \ref{fig2}a shows the spatial distribution of the 1-ns emission from the ammonia discharge operating under the reference conditions, measured at 80 ns after the beginning of the NT, so at $t_0 +$ 80 ns. Along the \emph{x}-axis (longitudinal direction), the pure ammonia plasma-emitted light extends beyond the dimensions of the copper electrodes and appears predominantly homogeneous (or diffuse), except for the edge effects. Along the \emph{y}-axis, the light intensity is distributed in three distinctive areas. From the upper to the lower dielectric, a thin luminous area is observed, followed by a darker space and then a large luminous zone. As shown with the axial emission distribution (along the \emph{y}-axis) plotted in figure \ref{fig2}b, the thinner luminous area intensity is greater than the larger luminous area. This pattern resembles that observed in a diode-type discharge. A similar analogy has already been drawn for some DBD-type discharges in sinusoidal mode, operating at pressures close to atmospheric pressure, with gases such as He, Ne, or gas mixtures referred to as Penning mixtures such as Ar:\ce{NH3} He:\ce{N2} \cite{hoft_exploring_2023, belinger_transition_2017, massines_recent_2009, yao_characteristics_2018, gherardi_transition_2000, bouzidi_maximization_2014, bazinette_optical_2015}. Thus, by analogy, the smallest and brightest luminous area would be attributed to the cathode fall and negative glow region, followed by a Faraday dark space and then a positive column close to the lower dielectric surface. Such an attribution implies that the surface of the upper dielectric assumes a cathodic role, while the lower dielectric acts as the anode during the negative transition. This assumption is consistent with the voltage-current measurements presented in our previous article \cite{Jean-Marie-Desiree_2024}, which indicate that the global electron flux $\Gamma_e$ is directed from the upper dielectric to the lower one during the negative transition.

\begin{figure}[h!]
	\centering
	\includegraphics[width=0.7\linewidth]{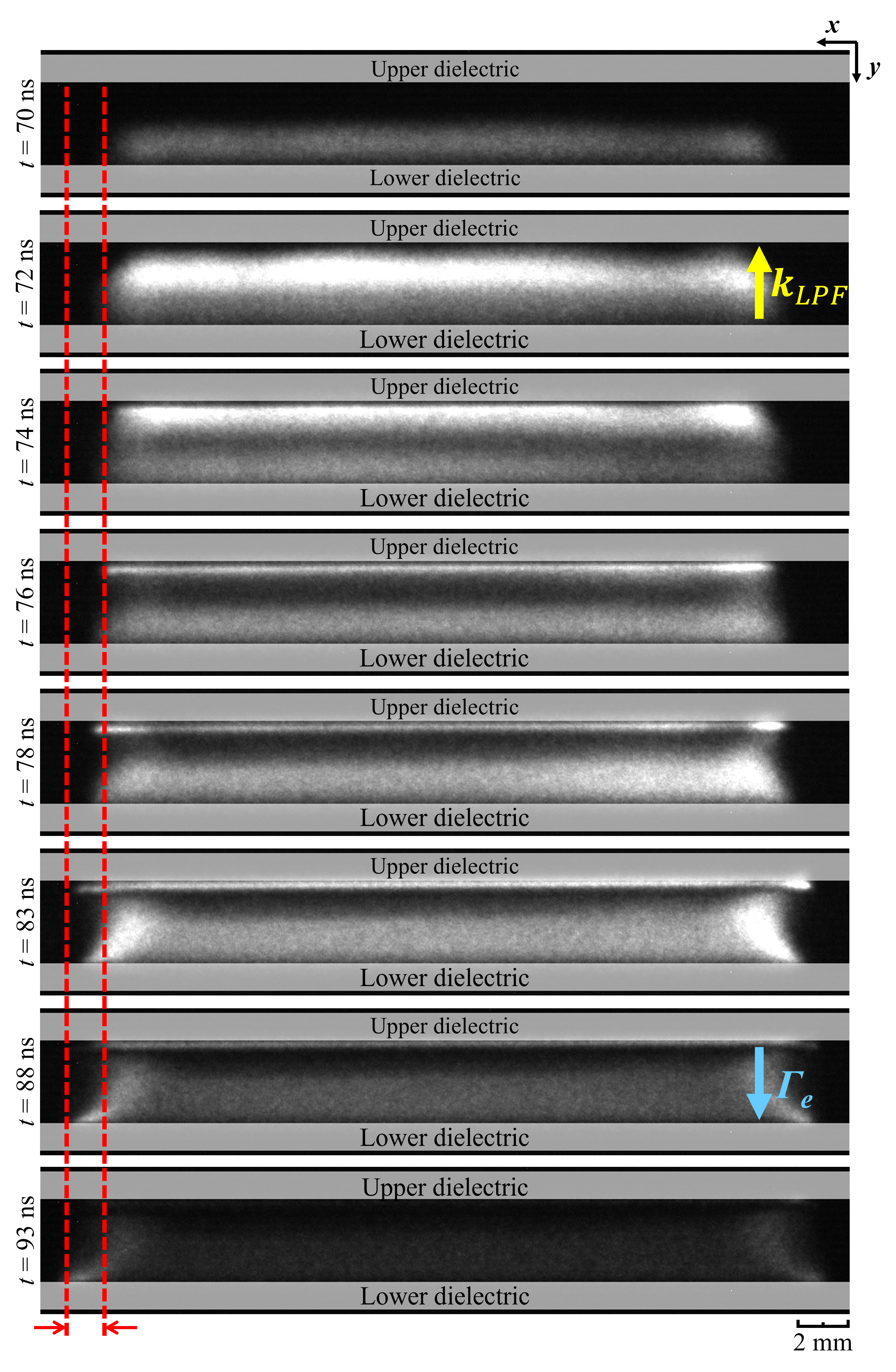}
	\caption{iCCD ultra-fast camera images of the iDBD discharge in pure ammonia at different times during its development in the NT, as indicated by the timestamps on the left. Each 1-mm thick dielectric is drawn to scale. The longitudinal expansion (\emph{x}-axis) is highlighted by the space outlined by the red dashed lines. Coloured arrows show the direction of the LPF in yellow and of the electron flux $\Gamma_e$ in blue. Measurements were taken at $10^4$ Pa, 8 kV$_{pp}$, and \emph{g} = 3 mm.}
	\label{fig3}
\end{figure}

The selected images compiled in figure \ref{fig3} depict the temporal evolution of the distribution of the discharge emitted-light, which establishes itself over the gas gap during the NT, under the reference conditions. This figure shows that the first variation in intensity is measured at $t_0$~+~70~ns, closest to the lower dielectric which acts as an anode from the start of the NT ($t_0$). These initial variations in intensity measured by the iCCD camera indicate the precedence of an excitation and de-excitation process that presumably occurs before $t_0$~+~70~ns. At $t_0$~+~72~ns, a region near the cathode becomes significantly brighter compared to the one observed at $t_0$~+~70~ns, which persists near the anode. Furthermore, these spatial variations in light intensity form a luminous propagation front (LPF), defined as the boundary of the brightest region located on the cathodic side. After $t_0$~+~74 ns, the LPF has finally reached the vicinity of the upper dielectric surface and can no longer be defined. Unlike $\Gamma_e$, this luminous front has thus propagated toward the cathode with an average velocity of at least 0.5 mm/ns (equivalent to $0.5 \times 10^6$ m/s). From this time, here $t_0$~+~74~ns, the gas gap can be divided into three distinct regions: the brightest region near the cathode carried by the LPF, a dark space, and a region near the anode with lower brightness. Once these three zones are established, a stationary regime will be spatially achieved along the \emph{y}-axis as shown with the following images in figure \ref{fig3}. Indeed, at $t_0$~+~76~ns, the previously described profile (cf. figure \ref{fig2}) is fully established, in accordance with the direction of $\Gamma_e$, as indicated by the blue arrow in figure \ref{fig3}. From $t_0$~+~76~ns onward, the distribution of the luminous and dark regions remains almost unchanged along the \emph{y}-axis.

However, an expansion of about 2.4 mm at the edge is then observed along the \emph{x}-axis under the reference conditions, as indicated by the dashed red lines in figure \ref{fig3}. Both sides of the iDBD cell are concerned. This longitudinal expansion is accompanied by an increase in light intensity at the edges of the discharge-emitted light due to the intrinsic limits of the copper electrodes (edge effects). Moreover, the shape of the extremities is altered, becoming more concave as the light dims. These observations seem to indicate the establishment of an electron gradient directly dependent on the evolution of the electric field. As soon as the electric field decreases in the iDBD cell, this then favours the diffusion of electrons along the longitudinal \emph{x}-axis.

\begin{figure}[h!]
	\centering
	\includegraphics[width=0.7\linewidth]{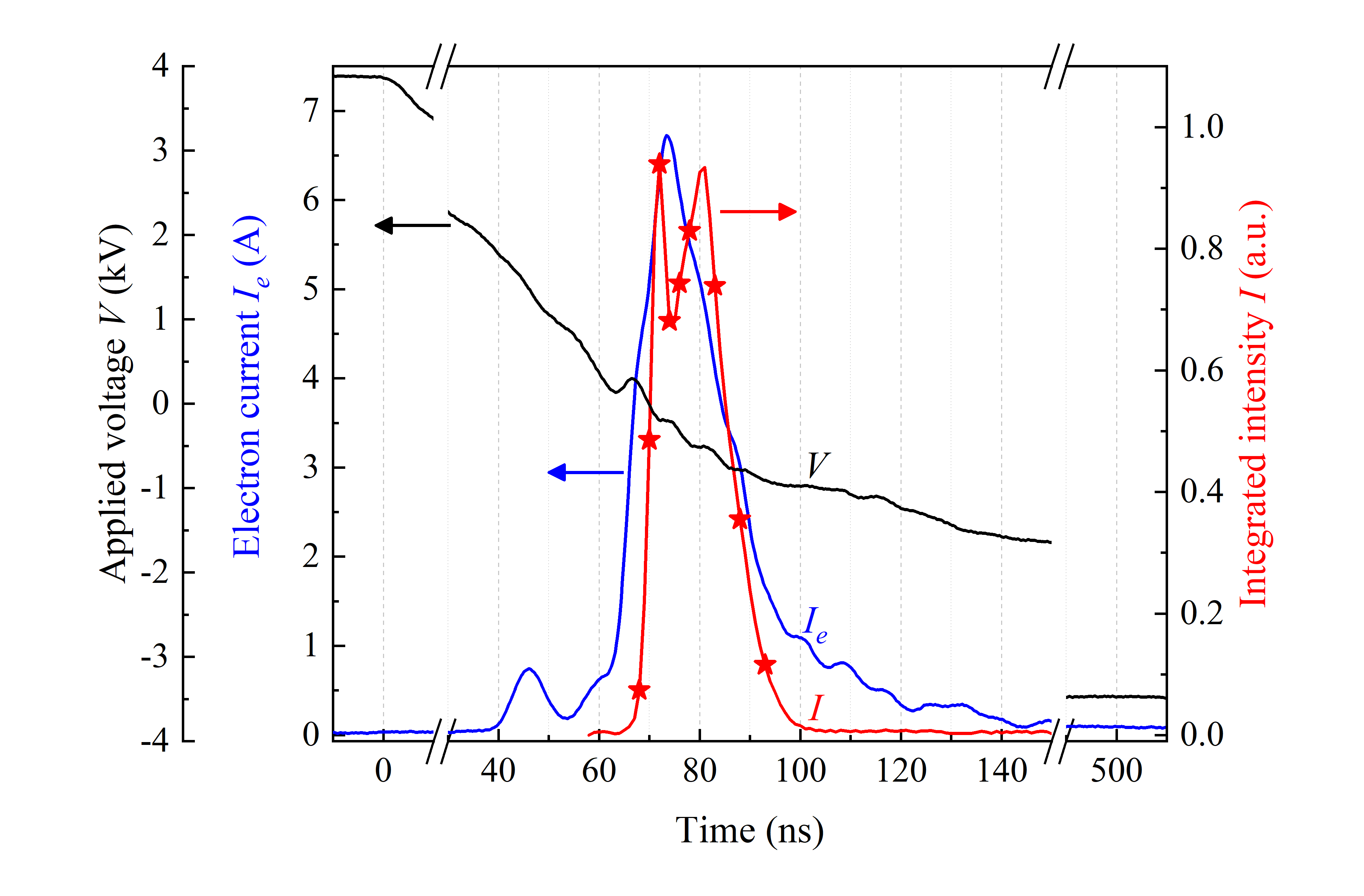}
	\caption{Evolution of the electrical caracteristic (applied voltage \emph{V} in black and electron current $I_e$ in blue) compared with that of the discharge emission intensity (in red) during NT. The discharge emission intensity was integrated over a 2 mm-wide region centred on the iDBD cell. The star symbols indicate the selected time points used to illustrate the emission dynamics in figure \ref{fig3}.}
	\label{fig4}
\end{figure}

By comparing the intensity integrated over a 2 mm-wide region centred on the iDBD cell with the discharge current ($I_{dis}$), a strong temporal correlation can be observed, as shown in figure \ref{fig4}. This correlation indicates that the light emission intensity closely mirrors the discharge current evolution, the latter thus describing the discharge establishment corresponding to the LPF phase during the current rise, and its subsequent extinction as the current decays. Moreover, the current discharge $I_{dis}$ that rises after $t_0$ + 60 ns is composed of both the displacement (sometimes referred as capacitive) current, indirectly linked to the voltage slope, and the conductive current, which describes the flow of charge through the inter-electrode space, as discussed in our previous article (cf. \cite{Jean-Marie-Desiree_2024}, figure 9). Under our reference conditions, $I_{dis}$ is mainly governed by the conductive current, while the displacement one appears to be negligible. More importantly, the temporal characteristics of this discharge current which include a rise time of about 10 ns and a duration of about 60 ns imply a short timescale over which ion motion can reasonably be considered slow or quasi-immobile compared with that of electrons. Consequently, the current $I_{dis}$ is predominantly due to electron dynamics and is therefore referred to as an electron current in our conditions. Taking this into account, the correlation between the electron current and the integrated discharge-emitted light over time highlights the dynamic coupling between charge transport and excitation processes, which seems to mainly originate from electron-impact mechanisms.

To better understand the formation of the luminous front, T. Hoder \emph{et al.} \cite{hoder_comparative_2010} tracked the spatio-temporal evolution of the emission lines of the \ce{N2} second positive system ($\lambda$ = 337.1 nm) and the \ce{N2} first negative system ($\lambda$ = 391.5 nm) in an atmospheric air discharge using a cross-correlation spectroscopy technique \cite{wagner_crosscorrelation_2005, Kozlov_2001}. T.~Hoder \emph{et al.} report an exponential increase in intensity near the anodic surface prior to the LPF phase, also referred to by the authors as a cathode-directed ionisation wave. The authors assumed that these variations in emission intensity serve as a marker of electron density variations, this exponential increase appears similar to the behaviour described by the Townsend avalanche mechanism. One of their proposed hypotheses is based on the concept of critical charge density ($n_{crit}$) required for the formation of the LPF phase. Thus, when the density of produced electrons $n_e$ becomes locally sufficient ($n_e$ > $n_{crit}$), the resulting space charge field would induce locally a distortion of the electric field and the modification of the electric potential distribution. Consequently, the initial ionisation and excitation volume must vary spatially to generate a displacement of the latter. This leads to the propagation of a front toward the cathode (LPF), which itself is governed by the temporal variation of the local electron density in the gap. In other words, the LPF phase would result from a "step-by-step" displacement driven by the expansion of the main electron production volume. Additionally, since our established iDBD discharge can be likened to a glow-type discharge (cf. figure \ref{fig2}) \cite{belinger_transition_2017, massines_recent_2009, bouzidi_maximization_2014, Bajon_2023}, the LPF phase would be analogue to the development and establishment of a cathode fall.

\begin{figure}[h!]
	\centering
	\includegraphics[width=0.7\linewidth]{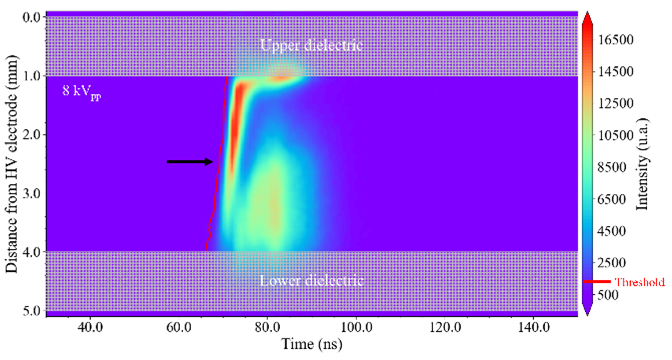}
	\caption{Temporal evolution of the axial profile of the light intensity distribution, integrated over 2 mm around the longitudinal centre of the iDBD cell through the gap. The vertical scale is referenced to the high-voltage (HV) copper electrode surface located on the top of the upper dielectric. A linear rainbow colour scale (from violet to red) is used to represent the intensity evolution. On this colour scale, the threshold level is set at 1.5 times the lowest value (500) to track the LPF position. Measurements were taken in pure \ce{NH3} at $10^4$ Pa with a 3-mm gap and 8 kV$_{pp}$.}
	\label{fig5}
\end{figure}

To study the LPF phase over time, the temporal evolution of the axial light intensity profile was integrated within a 2-mm region around the iDBD longitudinal centre. This process was applied to each 1-ns iCCD image based on figure \ref{fig3}, and the results are compiled into a two-dimensional representation, as shown in figure \ref{fig5}. Overall, figure \ref{fig5} summarises the temporal behaviours observed along the \emph{y}-axis discussed with figure \ref{fig3}. More precisely, figure \ref{fig5} illustrates the LPF phase which leads to the typical profile observed in a glow-type discharge. The positions of the dielectrics are indicated by two grey-hatched regions. Note that the variations in light intensity measured in these two areas result from reflections on the dielectric surfaces. To extract information related to the LPF phase, the images are analysed using a threshold condition (indicated in red on the colour scale). In other words, this threshold condition enables the detection of the initial variations in light intensity as a function of the distance from the high-voltage copper electrode and time, defined when the measured intensity exceeds 70\% of the background noise (500 a.u.). This approach results in the spatio-temporal tracking of the LPF position evolution. This threshold is used for all experimental conditions.

Since the behaviour of the LPF seems to be driven by ionisation mechanisms with limited spatial and temporal dimensions, the influence of parameters like voltage must be examined. The next subsection aims to explore the behaviour of the LPF phase as a function of the applied voltage.

\subsection{Influence of the applied voltage on the LPF phase}\label{voltage}

The study of the applied voltage influence has been conducted within the range of 4 to 8 kV$_{pp}$, while keeping other parameters constant (i.e., $10^4$ Pa and a 3-mm gas gap) as written in table \ref{table1}.

\begin{figure}[h!]
	\centering
	\includegraphics[width=0.7\linewidth]{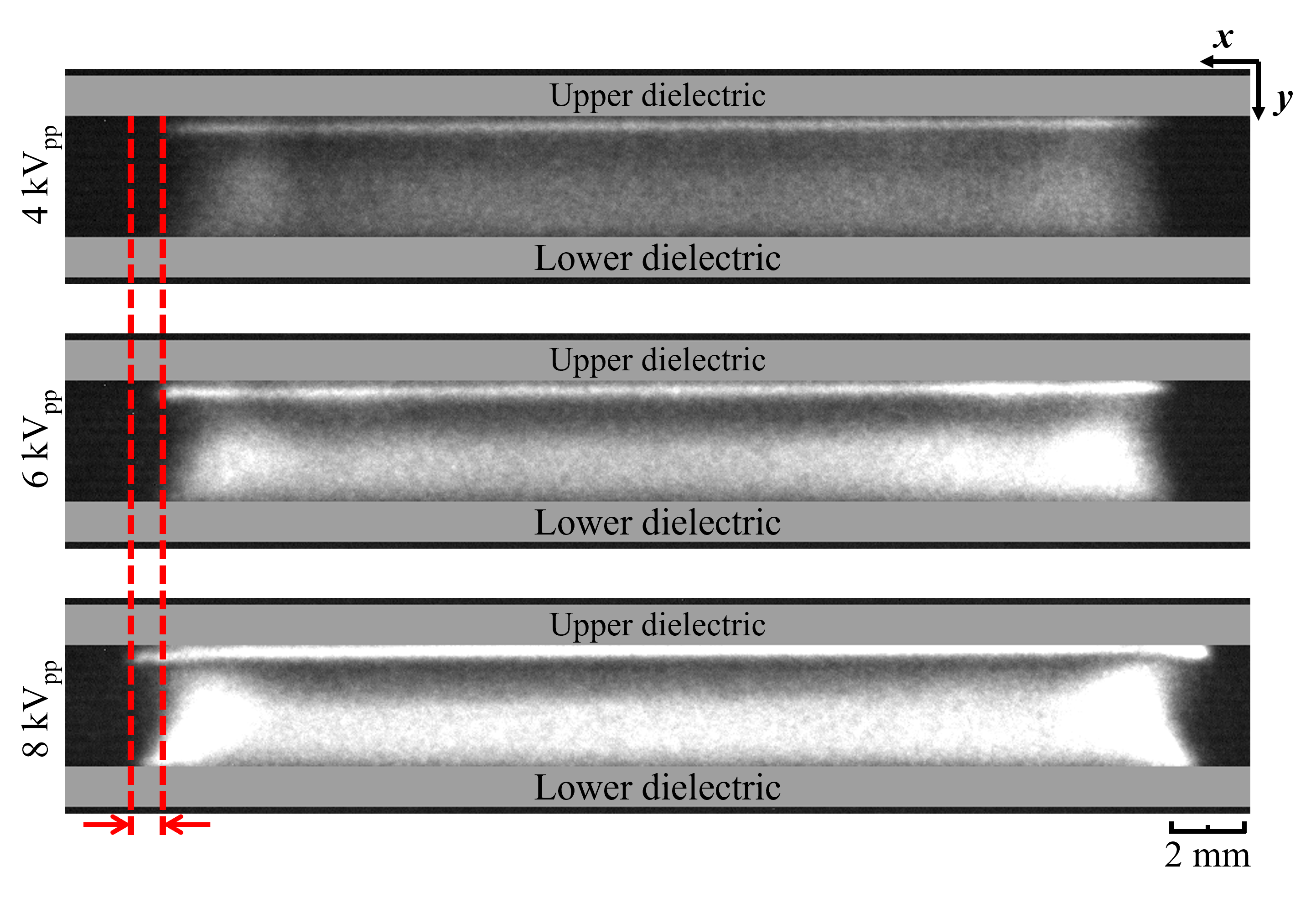}
	\caption{iCCD ultra-fast camera images of the iDBD discharge in pure ammonia after its establishment, for three applied voltage conditions as indicated on the left. The longitudinal increase is highlighted by the space outlined by the red dashed lines. Measurements were taken at $10^4$ Pa and \emph{g} = 3 mm.}
	\label{fig6}
\end{figure}

Before discussing the relationship between the LPF phase and the electrical characteristics, figure \ref{fig6} presents the spatial distribution of the 1-ns discharge-emitted light in the gas gap after a stationary regime has been established along the \emph{y}-axis ($t_0$ + 124, 86 and 83 ns for 4, 6 and 8 kV$_{pp}$, respectively). Compared to the reference conditions discussed previously, the 4 and 6 kV$_{pp}$ conditions exhibit a similar distribution of luminous and dark regions, corresponding to a glow-type discharge. However, the longitudinal dimension of the discharge observed along the \emph{x}-axis is larger as \emph{V} increases. Indeed, the dashed red lines drawn in figure \ref{fig6} highlight an increase of about 1.2 mm, between the 4 and 8 kV$_{pp}$. This longitudinal increase in volume with the increase in \emph{V} seems to be linked to the increase in the measured discharge current $I_{dis}$ and thus to the charge density $n_c$ in the volume \cite{Jean-Marie-Desiree_2024}. This increase in $n_c$ would amplify the $n_e$ gradient from the edges towards the centre of the discharge. The result would be a greater electrostatic force which is hypothetically towards the edges of the discharge, driving more electrons in that direction as $n_c$ increases. Moreover, the overall luminous intensity in the inter-dielectric space increases by a factor of 2.4 between 4 and 8 kV$_{pp}$. These first observations suggest that discharge-emitted light distribution is largely determined by the electrical characteristics and therefore likely influenced by spatio-temporal phenomena such as the LPF.

\begin{figure}[!htb]
	\begin{minipage}{0.48\textwidth}
		\centering
		\includegraphics[width=\linewidth]{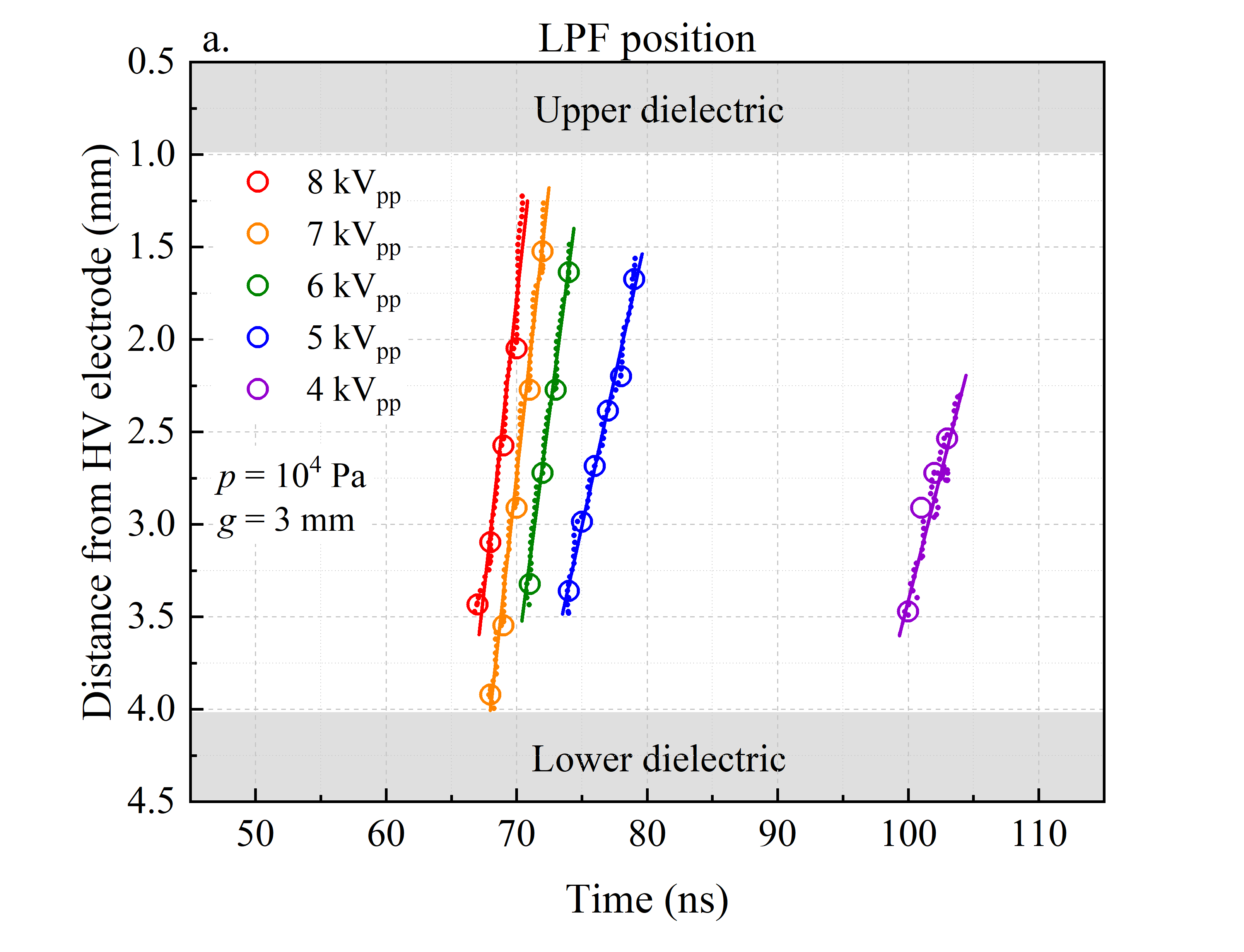}
	\end{minipage}\hfill
	\begin{minipage}{0.48\textwidth}
		\centering
		\includegraphics[width=\linewidth]{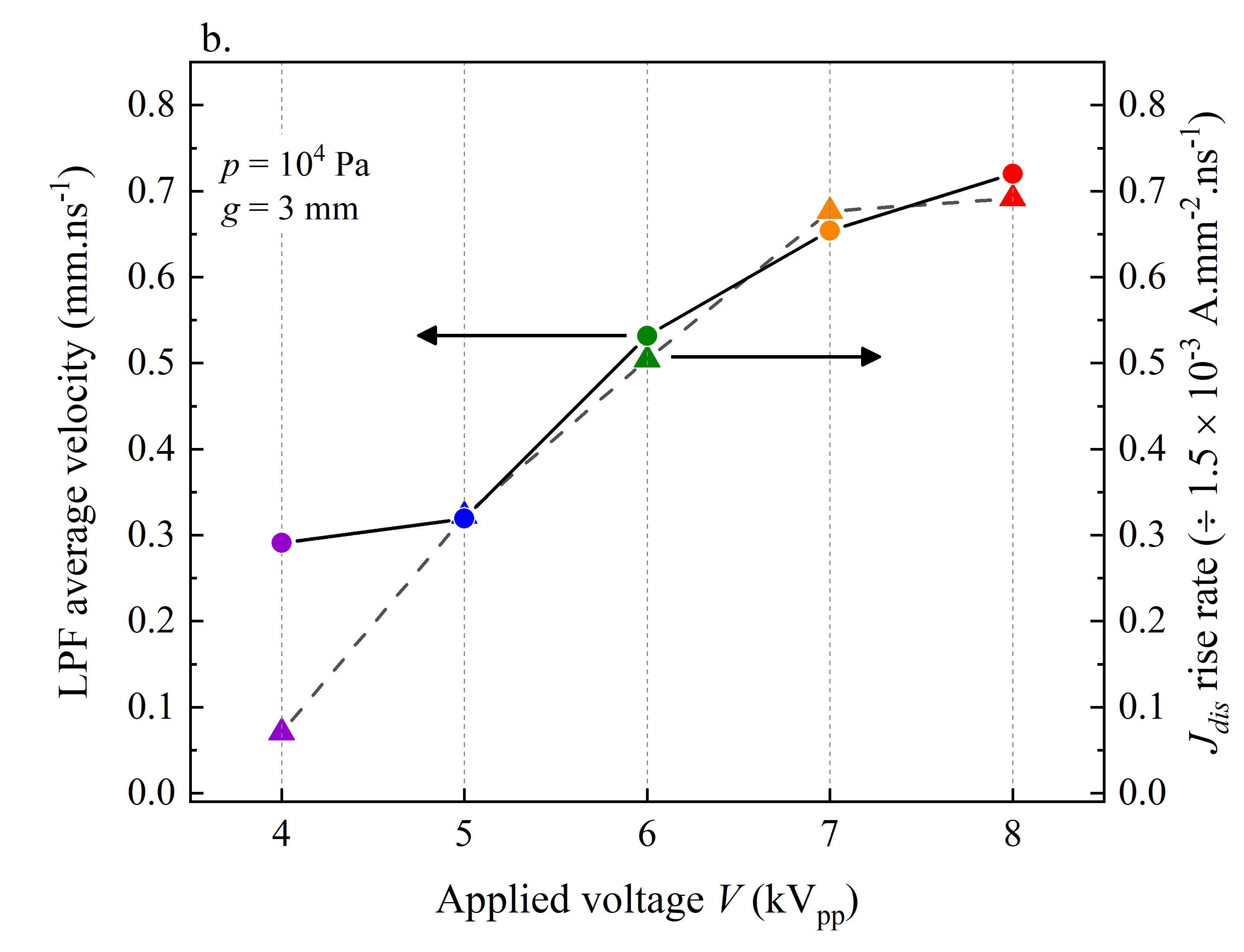}
	\end{minipage}
	\caption{(a) Evolution of the LPF position in the gap for five applied voltage conditions during the NT. The experimental data are represented by hollow symbols. The interpolation of these data is shown as a scatter plot, while their linear fit is depicted by solid lines. The slope values allow for the determination of the LPF velocity, which is then plotted as circular symbols in (b). This LPF velocity is compared with the rise rate of the surface current density $J_{dis}$, represented by triangular symbols and adjusted by a proportionality factor of 1.5 $\times ~10^{-3}$. Measurements were conducted in pure \ce{NH3} at $10^4$ Pa and \emph{g} = 3 mm.}
	\label{fig7}
\end{figure}

In order to study the spatial and temporal development of the LPF phase, this phenomenon has been tracked using the 1-ns iCCD images. The position of the LPF was then reported in figure \ref{fig7}a for the five studied applied voltage conditions. Figure \ref{fig7}a shows that the LPF position varies linearly over time, regardless of the applied voltage under our conditions. In addition, the average velocity calculated from the time derivative of these curves indicates for example a faster LPF phase at 8 kV$_{pp}$ compared to the 5 kV$_{pp}$ condition, with average velocities of approximately 0.70 mm/ns (0.7 $\times ~10^6$ m/s) and 0.3 mm/ns (0.3 $\times ~10^6$ m/s) respectively, as shown in figure \ref{fig7}b.

Assuming that the LPF phase results from a critical charge density being reached, which occurs more rapidly at higher voltage, a correlation between the LPF velocity and the establishment of the discharge current $I_{dis}$ should be identified. Indeed, the LPF phase occurs over a time interval during which $I_{dis}$ increases linearly. This discharge current $I_{dis}$ has already been plotted for each of the studied conditions and discussed in our previous article \cite{Jean-Marie-Desiree_2024}. To establish this correlation in a broader context and by considering that the plasma discharge is in diffuse mode (homogeneous), the surface current density $J_{dis}$ should be considered. The latter is defined as the ratio of $I_{dis}$ to the apparent surface area of the copper electrodes. Therefore, the time derivative of $J_{dis}$ has been calculated over the same time interval as the LPF phase occurs for each voltage condition.

Thus, the results have been plotted and corrected by a factor of 1.5 $\times$~$10^{-3}$ in figure \ref{fig7}b. These results clearly show that the dynamics of the LPF are correlated with the current measurements which account for all the charge production and consumption phenomena occurring within the volume. A slight deviation in the correlation is observed under the 4 kV$_{pp}$ condition, which lies near the discharge breakdown threshold and leads to an unstable initial and post-discharge state between successive discharges. Based on this correlation, it is reasonable to assume that the LPF velocity is closely related to the temporal variation of the volumetric charge density, which both increase with the applied voltage \emph{V}. Indeed, the latter influences the ionisation and excitation rates of the gas species. In other words, a higher applied voltage induces in our case a greater $\frac{dV}{dt}$ (see \cite{Jean-Marie-Desiree_2024}) and implies a higher instantaneous charge density in the volume. As a result, the critical charge density $n_{crit}$ is reached more quickly at higher voltages. If this hypothesis regarding the existence of a $n_{crit}$ value may be verified, it also implies that a faster deformation of the equipotential lines ahead of the LPF leads to a highly localised and more rapid charge displacement. The displacement mechanism proposed here follows a "step-by-step" propagation process towards the cathode. The results presented in figure \ref{fig7}b support the hypothesis that variations in charge density within the inter-dielectric space are a necessary condition for the development of the LPF phase. In other words, if there is no temporal variation in charge density exceeding the $n_{crit}$ value, the LPF phase cannot occur. Furthermore, this hypothesis also suggests that the LPF delineates the moving region where most of the charged particles are generated, enabling the $n_{crit}$ to be reached at a given point in space. Expressed differently, this corresponds to the definition of an ionisation front/wave, as suggested in the literature \cite{hoder_comparative_2010, Vasilyak_1994}.

In summary, the LPF phase appears to correspond to the displacement of the electron production in the gas volume, ultimately leading to the establishment of a glow discharge. Since the "step-by-step" displacement is triggered when the $n_{crit}$ is reached at a given spatial point, the LPF is faster as \emph{V} and $\frac{dV}{dt}$ increase. Therefore, this excitation and ionisation wave propagates along the electric field established in the inter-dielectric space, shaped by the previous discharge occurring during the PT (see \cite{Jean-Marie-Desiree_2024}). In other words, this propagation can only occur in the direction opposite to $\Gamma_e$, that is towards the cathode under our conditions.

If this correlation between the LPF velocity and the $J_{dis}$ rise rate seems to be valid, the influence of the pressure must be also studied. Indeed, an increase in gas particle density leads to electron cooling due to collisions, which, in turn, would reduce the LPF velocity.

\subsection{Influence of the ammonia pressure on the LPF phase}\label{pressure}

Following the same methodology as in the previous subsection, the influence of the ammonia pressure was studied within the range of 0.8 to 2.0 $\times$~$10^4$ Pa, while keeping other parameters constant (i.e., 6 kV$_{pp}$ and a 3-mm gas gap), as written in table \ref{table1}.

\begin{figure}[h!]
	\centering
	\includegraphics[width=0.7\linewidth]{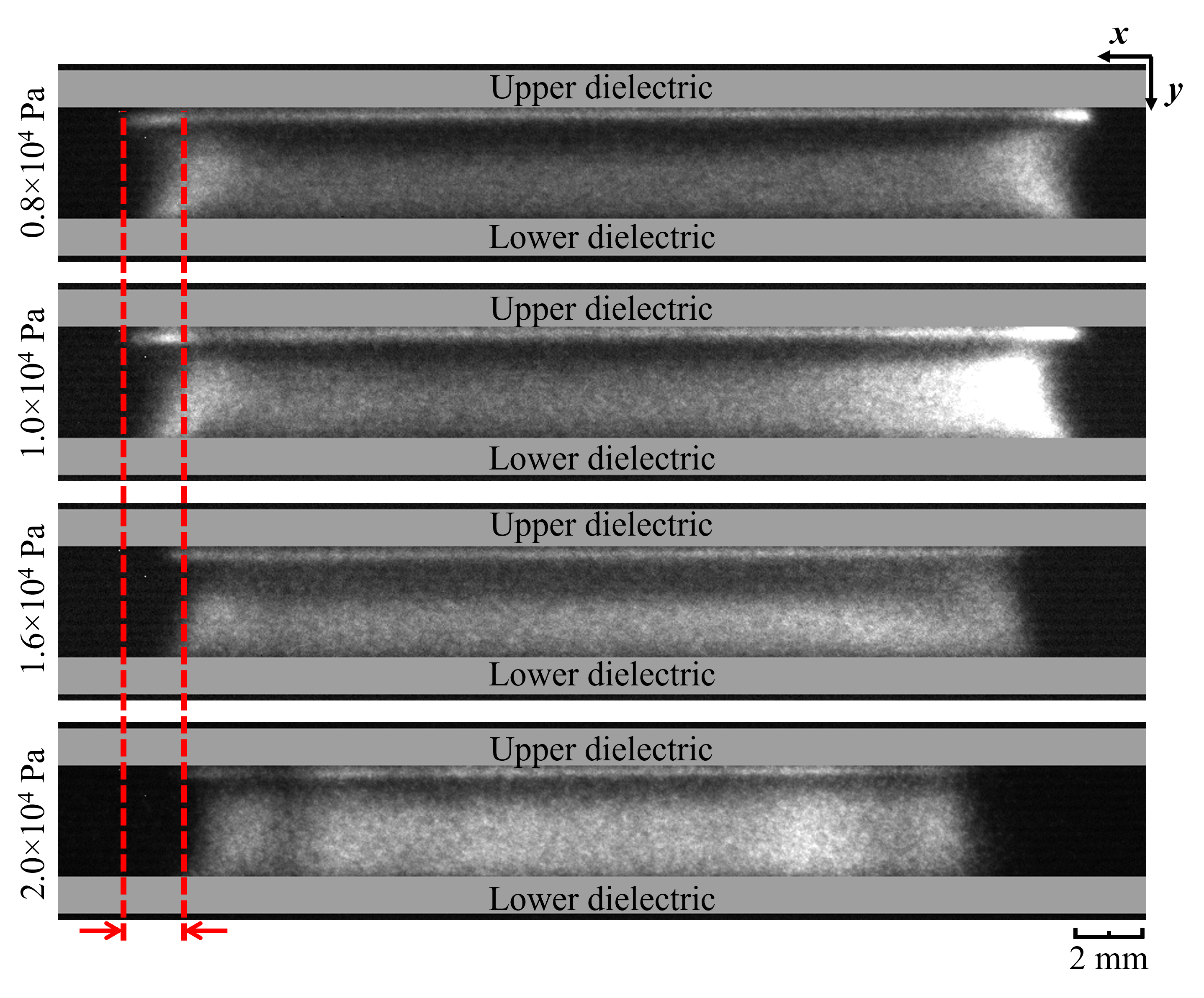}
	\caption{iCCD ultra-fast camera images of the iDBD discharge in pure ammonia after its establishment, for four pressure conditions as indicated on the left. The longitudinal increase is highlighted by the space outlined by the red dashed lines. Measurements were taken at 6 kV$_{pp}$ and \emph{g} = 3 mm.}
	\label{fig8}
\end{figure}

Figure \ref{fig8} presents the spatial distribution of the 1-ns discharge-emitted light for four pressure conditions: 0.8, 1.0, 1.6 et 2.0 $\times$~$10^4$ Pa at $t_0$ + 80, 87, 97 and 128 ns, respectively. Unlike in figure \ref{fig6}, the grayscale distribution in figure \ref{fig8} could not be kept constant for all images, though it is worth noting that the same camera parameters and the threshold condition were maintained. Indeed, the measured light intensity decreases significantly under our conditions with increasing ammonia pressure.

This trend is obviously correlated with the decrease in $I_{dis}$ amplitude observed \cite{Jean-Marie-Desiree_2024}, as the increase in pressure is unfavourable to efficient ionisation and excitation mechanisms in this range of pressures (due to significant cooling of electrons by collisions). Furthermore, in comparison with the reference conditions discussed in subsection \ref{gen_obvs}, the 0.8 to 1.6 $\times$~$10^4$ Pa conditions exhibit similarities to a glow-type discharge, from a phenomenological perspective. However, the image for the 2.0 $\times$~$10^4$ Pa condition shows a distinct behaviour with the presence of slight over-intensities, excluding edge effects, giving a columnar-like aspect to the discharge. This discharge regime is referred to as the columnar (or self-organised) mode \cite{Jean-Marie-Desiree_2024, radu_2003}. Additionally, the longitudinal dimension of the discharge decreases as the pressure increases from 0.8 to 2.0~$\times$~$10^4$, by more than 4.5 mm, as indicated by the red dashed lines in figure \ref{fig8}. Indeed, as the gas density increases, the mean free path of particles necessarily decreases, limiting the transport processes along this \emph{x}-axis. This is further compounded by the decrease in $I_{dis}$ with gas pressure which also contributes to modify the electron density gradient and to limit the electron diffusion due to collision processes in the gas volume.

In addition to the longitudinal shrinkage, the columnar-like aspect observed beyond 1.6 $\times$~$10^4$ Pa indicates that under these conditions, the discharge spatially reorganises with the increase in gas particle density \cite{kogelschatz_filamentary_2002, yao_characteristics_2018, raizer_physical_2013}. In comparison, C. Yao \emph{et al.} \cite{yao_characteristics_2018} report a self-organisation behaviour in their study of a sinusoidal DBD discharge with an Ar:\ce{NH3} gas mixture at atmospheric pressure. Using spectroscopic analysis coupled with simulation results, the authors propose a hypothesis for the emergence of this columnar mode based on a competition between production (e.g., Penning ionisation) and loss reactions (e.g., electron attachment) of free electrons as a function of the \ce{NH3} volumetric fraction. According to the authors, this competition highlights that insufficient electron density and temperature lead to an auto-organisation of the discharge. In our case, this seems to be mainly due to the increase in pure ammonia gas pressure, i.e., to the decrease of the diffusion process of electrons in the volume.

\begin{figure}[!htb]
	\begin{minipage}{0.48\textwidth}
		\centering
		\includegraphics[width=\linewidth]{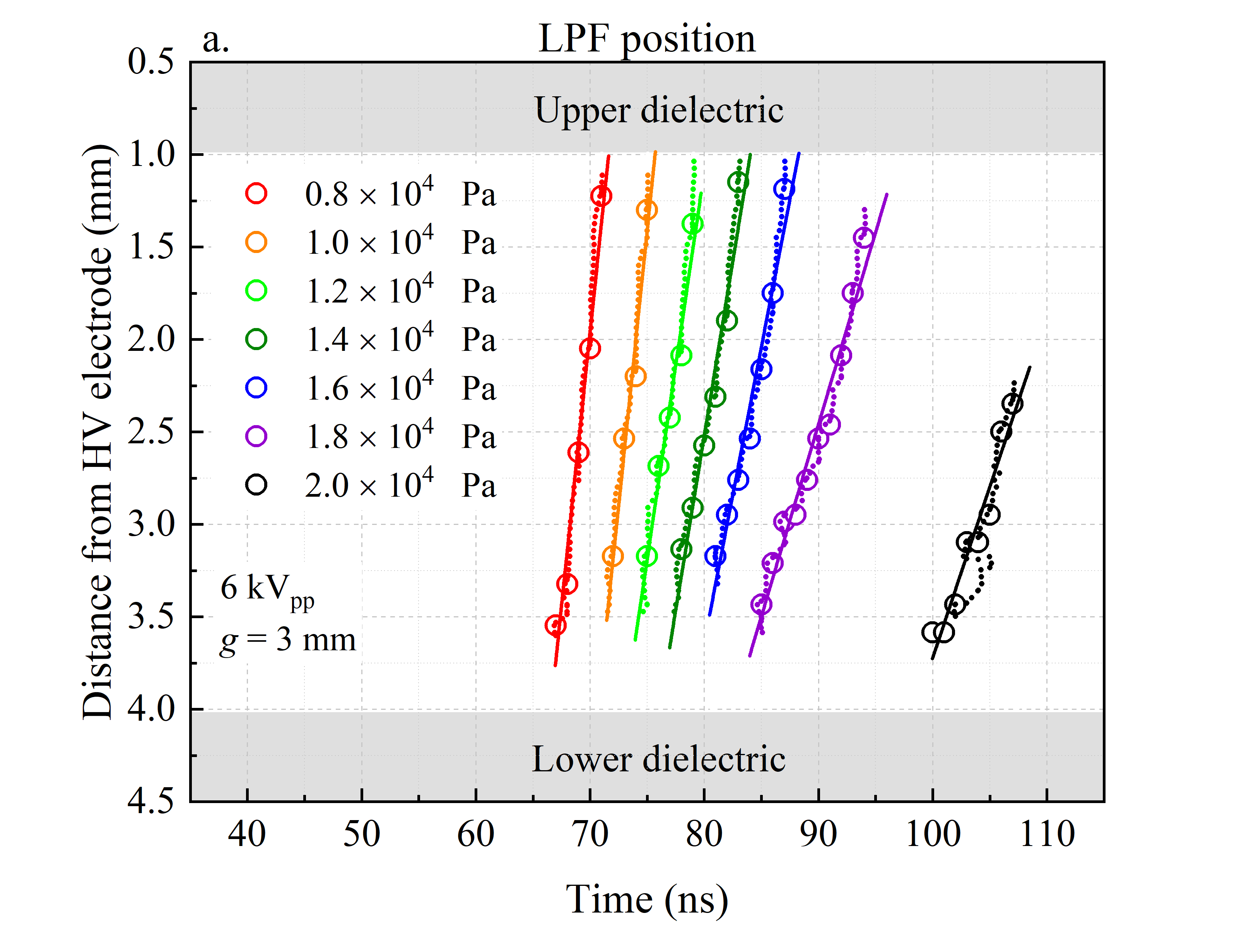}
	\end{minipage}\hfill
	\begin{minipage}{0.48\textwidth}
		\centering
		\includegraphics[width=\linewidth]{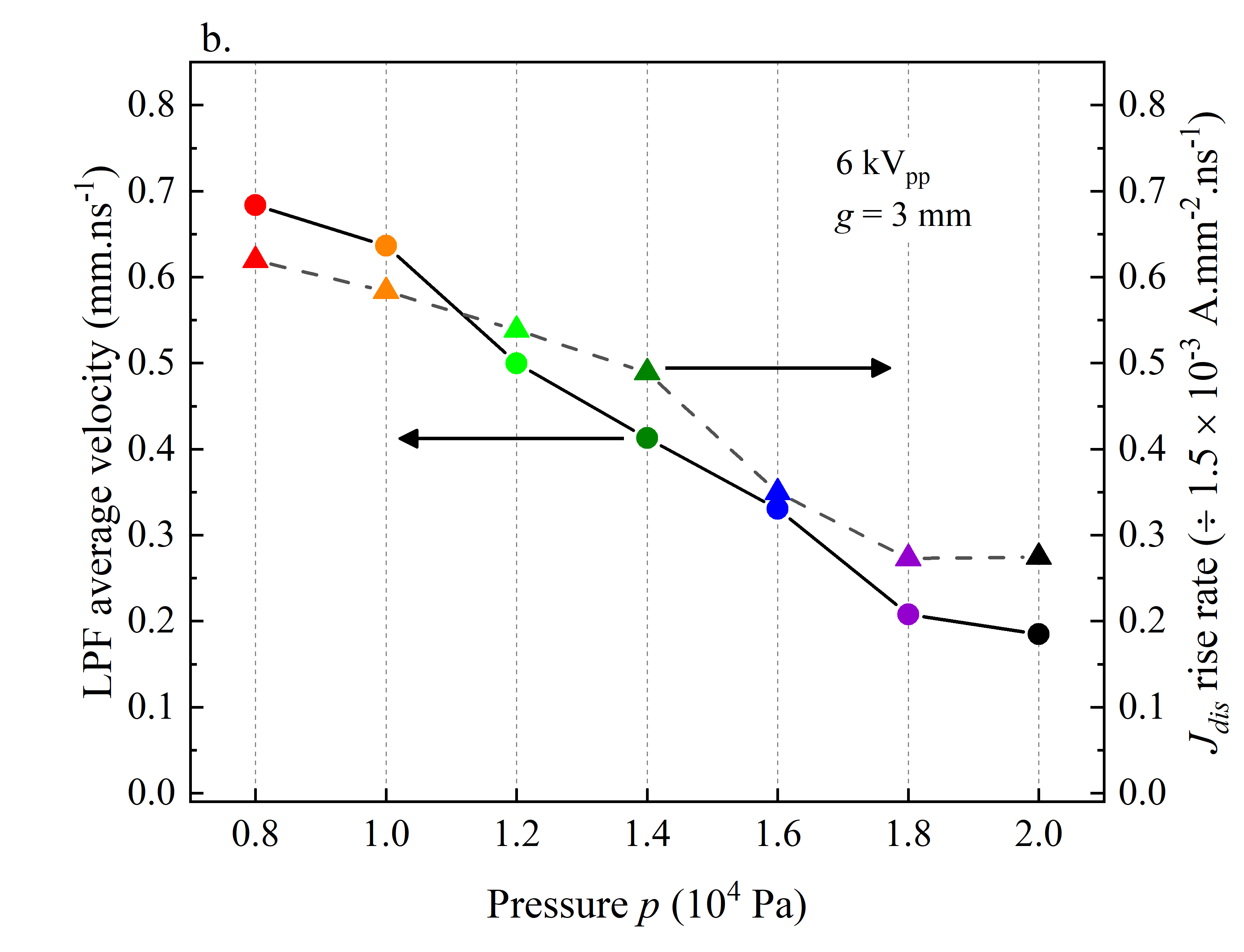}
	\end{minipage}
	\caption{(a) Evolution of the LPF position in the gap for seven pressure conditions during the NT. The experimental data are represented by hollow symbols. The interpolation of these data is shown as a scatter plot, while their linear fit is depicted by solid lines. The slope values allow for the determination of the LPF velocity, which is then plotted as circular symbols in (b). This LPF velocity is compared with the rise rate of the surface current density $J_{dis}$, represented by triangular symbols and adjusted again by a proportionality factor of 1.5 $\times$~$10^{-3}$. Measurements were conducted in pure \ce{NH3} under 6 kV$_{pp}$ and \emph{g} = 3 mm.}
	\label{fig9}
\end{figure}

To build upon these observations, the following discussion explores the relationship between the average LPF velocity and the rise rate of $J_{dis}$. Figure \ref{fig9} is plotted using the same methodology as previously mentioned (c.f. figure \ref{fig7}) but applied to different pressure conditions. Once more, the temporal evolution of the LPF plotted in figure \ref{fig9}a seems quite linear, regardless of the ammonia pressure within the studied range. Furthermore, the LPF average velocity decreases with pressure, for example by a factor of 3.5, from 0.7 to 0.2 mm.ns$^{-1}$ (or $\times$~$10^6$ m.s$^{-1}$) corresponding to 0.8 and 2.0 $\times$~$10^4$ Pa respectively, as shown in figure \ref{fig9}b.

In addition and as previously observed, the calculated LPF average velocities are in good agreement with the rise rate of $J_{dis}$, once again adjusted by a proportionality factor of 1.5 $\times$~$10^{-3}$, as shown in figure \ref{fig9}b. To better understand the effect of pressure while taking into account collisions within the volume, the evolution of the reduced electric field $\frac{E}{N}$ \cite{Vasilyak_1994, starikovskaia_pulsed_2001, anikin_polarity_2002} is relevant to examine. In our previous study \cite{Jean-Marie-Desiree_2024}, the electric field measurements performed in similar conditions show that at fixed applied voltage, as pressure increases, $\frac{E}{N}$ decreases, leading to a lower ionisation frequency and consequently a reduction in the local charge density that can be generated within the same volume and time frame. Since charge production is lower, the local critical charge density $n_{crit}$ is necessarily reached later at higher pressures (figure \ref{fig9}a) and the LPF velocity consequently decreases (figure \ref{fig9}b). The reduced mean free path contributes to greater energy dissipation within the system. As a result, the ionisation region contracts along the \emph{x}-axis. Overall, the LPF, which is associated with an ionisation and excitation volume as previously assumed, propagates more slowly. This further supports the "step-by-step" propagation mechanism of the LPF, which ultimately depends on the ionisation volume in which $n_{crit}$ can be reached.

The correlation between the LPF velocity and the $J_{dis}$ rise rate is thus verified for two operating parameters: the applied voltage \emph{V} and the pressure \emph{p} within our studied conditions. To definitely validate this correlation in diffuse mode, the next subsection deals with the influence of the gas gap \emph{g} on the iDBD discharge.

\subsection{Influence of the gas gap on the LPF phase}\label{gas_gap}

With the same methodology, the influence of the gas \emph{g} was studied within the range between 3 and 9 mm, while keeping other parameters constant (i.e., 6 kV$_{pp}$ and $10^4$ Pa), as written in table \ref{table1}.

\begin{figure}[h!]
	\centering
	\includegraphics[width=0.7\linewidth]{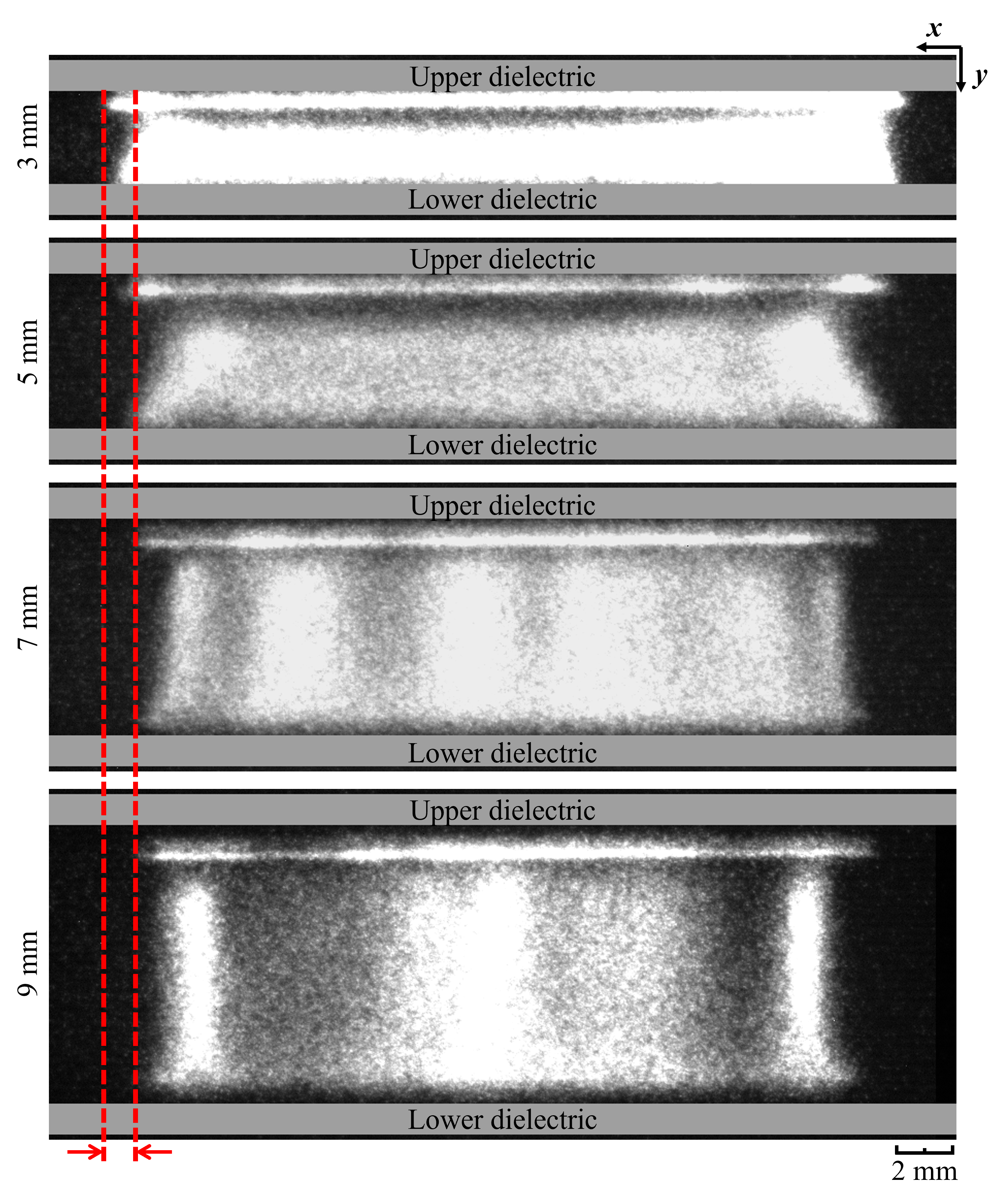}
	\caption{iCCD ultra-fast camera images of the iDBD discharge in pure ammonia after its establishment, for four gas gap \emph{g} conditions as indicated on the left. The longitudinal increase is highlighted by the gap outlined by the red dashed lines. Measurements were taken at 6 kV$_{pp}$ and $10^4$ Pa.}
	\label{fig10}
\end{figure}

Figure \ref{fig10} presents the spatial distribution of the 1-ns discharge-emitted light for four gas gap conditions: 3, 5, 7 and 9 mm at $t_0$ + 88, 98, 116 and 136 ns, respectively. Similarly to figure \ref{fig8}, the grayscale distribution could not be kept constant in figure \ref{fig10}. Note that in this figure, some frames appear intentionally overexposed but not saturated to capture the full dynamic range of the LPF emission while maintaining fixed camera gain and optimal signal-to-noise ratio across all conditions. However, since the camera settings remained unchanged, the measured light intensity decreases with the gas gap \emph{g}. This evolution is as expected correlated with the variation in the $I_{dis}$ amplitude \cite{Jean-Marie-Desiree_2024}. Indeed, the electric measurement from our previous study \cite{Jean-Marie-Desiree_2024} show that, at fixed voltage, the increase in \emph{g} weakens the established electric field in the gas gap, leading to lowering the ionisation and excitation rate. From the compiled images in figure \ref{fig10}, two distinct categories emerge.

First, the images from figure \ref{fig10} corresponding to the 3- and 5-mm gap conditions exhibit similarities in the diffuse behaviour of the discharge-emitted light, resembling the phenomenological characteristics of a glow-type discharge. Second, for the 7- and 9-mm gaps, the light emission adopts a structured columnar-like aspect. Thus, under these conditions, the 7-mm gap condition marks a transition from diffuse to columnar mode. 

\begin{figure}[!htb]
	\begin{minipage}{0.48\textwidth}
		\centering
		\includegraphics[width=\linewidth]{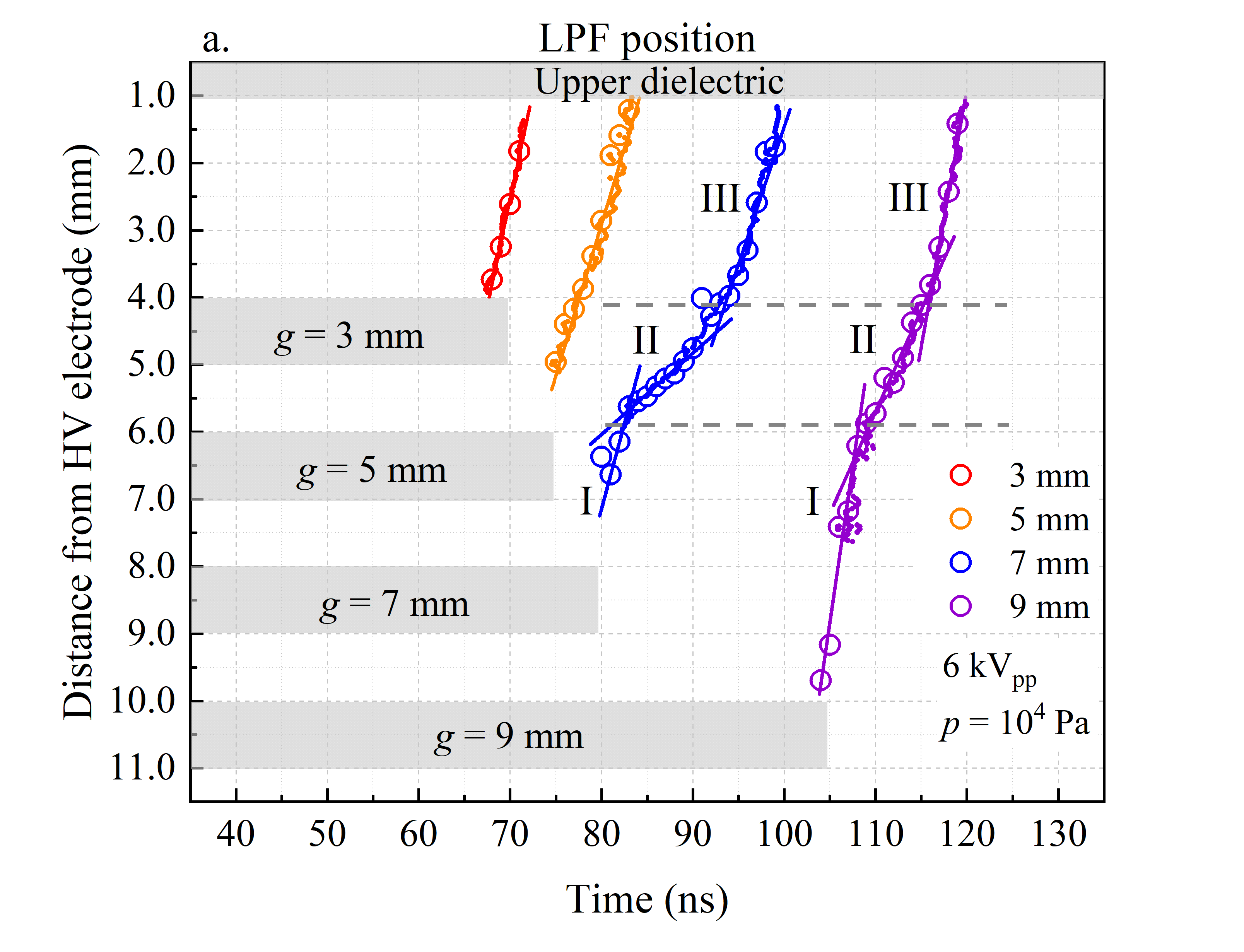}
	\end{minipage}\hfill
	\begin{minipage}{0.48\textwidth}
		\centering
		\includegraphics[width=\linewidth]{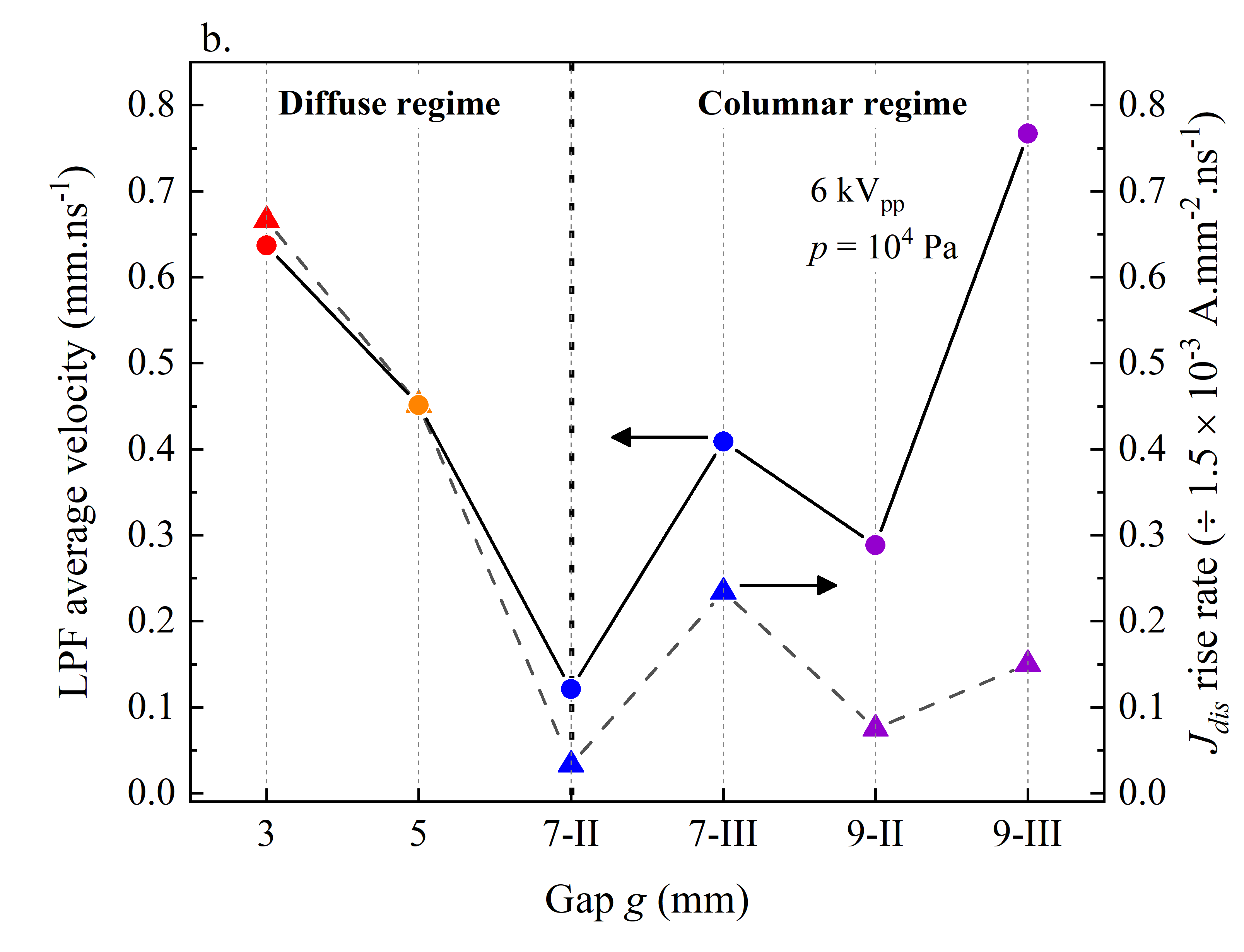}
	\end{minipage}
	\caption{(a) Evolution of the LPF position in the gap for five applied voltage conditions during the NT. The experimental data are represented by hollow symbols. The interpolation of these data is shown as a scatter plot, while their linear fit is depicted by solid lines. The dashed gray lines mark the slope changes observed for the 7- and 9-mm conditions, distinguishing three spatiotemporal domains annotated here with Roman numerals (I, II, III). The slope values allow for the determination of the LPF velocity, which is then plotted as circular symbols in (b). This LPF velocity is compared with the rise rate of the surface current density $J_{dis}$, represented by triangular symbols and again adjusted by a proportionality factor of $1.5 \times$~$10^{-3}$. Measurements were conducted in pure \ce{NH3} at $10^4$ Pa and under 6 kV$_{pp}$.}
	\label{fig11}
\end{figure}

Additionally, it is worth noting that the blurred appearance of these columns results from the depth of field effect of the iCCD camera optics (c.f. section \ref{experimental_setup}). In other words, this blur arises from the spatial integration of luminous columns developing outside the camera’s focal plane, rather than a stochastic occurrence of these columns from one discharge to another. In fact, one can note that these columns are remarkably reproducible from one period to the next.

Finally, figure \ref{fig10} also shows a decrease in the longitudinal dimension of the discharge by more than 3~mm in total for a threefold increase in \emph{g}. This result echoes the discussion on the presence of an electron density gradient within the discharge along the \emph{x}-axis. Based on all the spatial differences observed in the discharge structure over NT as a function of \emph{g}, the relationship between the LPF mean velocity and the $J_{dis}$ rise rate needs to be investigated.

\newpage
Thus, figure \ref{fig11}a provides a more detailed analysis of the temporal evolution of the LPF position as a function of the gas gap \emph{g}. While the experimental data for the 3- and 
5-mm conditions exhibit nearly linear variations of the LPF propagation, those for 7 and 9 mm reveal three distinct slopes of the LPF propagation, labelled as I, II, and III on figure \ref{fig11}a. As in the previous subsections dealing with the applied voltage \emph{V} and the pressure \emph{p} influences, a comparison has been made between these slopes and $J_{dis}$ rise rate, as shown in figure \ref{fig11}b. Beyond the slopes obtained for the 3- and 5-mm gaps, only slopes II and III from the 7- and 9-mm gap conditions could be compared to the $J_{dis}$ rise rate. Indeed, slopes labelled I correspond to the very early moments of the discharge (a few nanoseconds), where the very low values of $I_{dis}$  prevent an accurate determination of its temporal derivative. Nevertheless, for the 3- and 5-mm gaps, the LPF average velocity and the $J_{dis}$ rise rate are well correlated with a factor of again $1.5 \times$~$10^{-3}$. This correlation further reinforces the idea of a direct link between the LPF dynamics and the temporal evolution of $J_{dis}$, as previously observed in studies on voltage and pressure.

However, from the 7-mm case, a significant discrepancy emerges between the values associated with LPF and $J_{dis}$. This could be attributed to the transition to a columnar mode, as pictured in figure \ref{fig10} and also observed in previous section (in figure \ref{fig9}b, beyond $1.8 \times$~$10^4$ Pa condition). This discrepancy between the LPF velocity and $J_{dis}$ rise rate in columnar discharge modes may result from the reduction in the surface and volume interaction of the discharge with its surrounding media. In this specific case, $J_{dis}$ may be underestimated since its calculation is based on the total surface of the copper electrodes, while accurately estimating the interaction surfaces in the columnar mode remains challenging. A possible approach to addressing this issue could involve performing electrical measurements using segmented electrodes, a diagnostic technique developed by N. Naudé and coworkers \cite{tyl_new_2021} coupled with the use of transparent materials.

Based on these observations, a physical process is proposed for each slope. The first slope (I) corresponds to the initial formation of the luminous region near the anode, as previously mentioned. The second (II) and third (III) slopes correspond to the actual propagation of the LPF. More precisely, slope II relates to the gradual increase of the persistent luminous region near the anode over time, without causing its overall displacement towards the cathode. In contrast, slope III corresponds to the propagation of the most intense ionisation and excitation front, which detaches from the persistent region near the anode formed during phase II, leaving a dark space behind and accelerates towards the cathode.

\begin{figure}
	\centering
	\includegraphics[width=\linewidth]{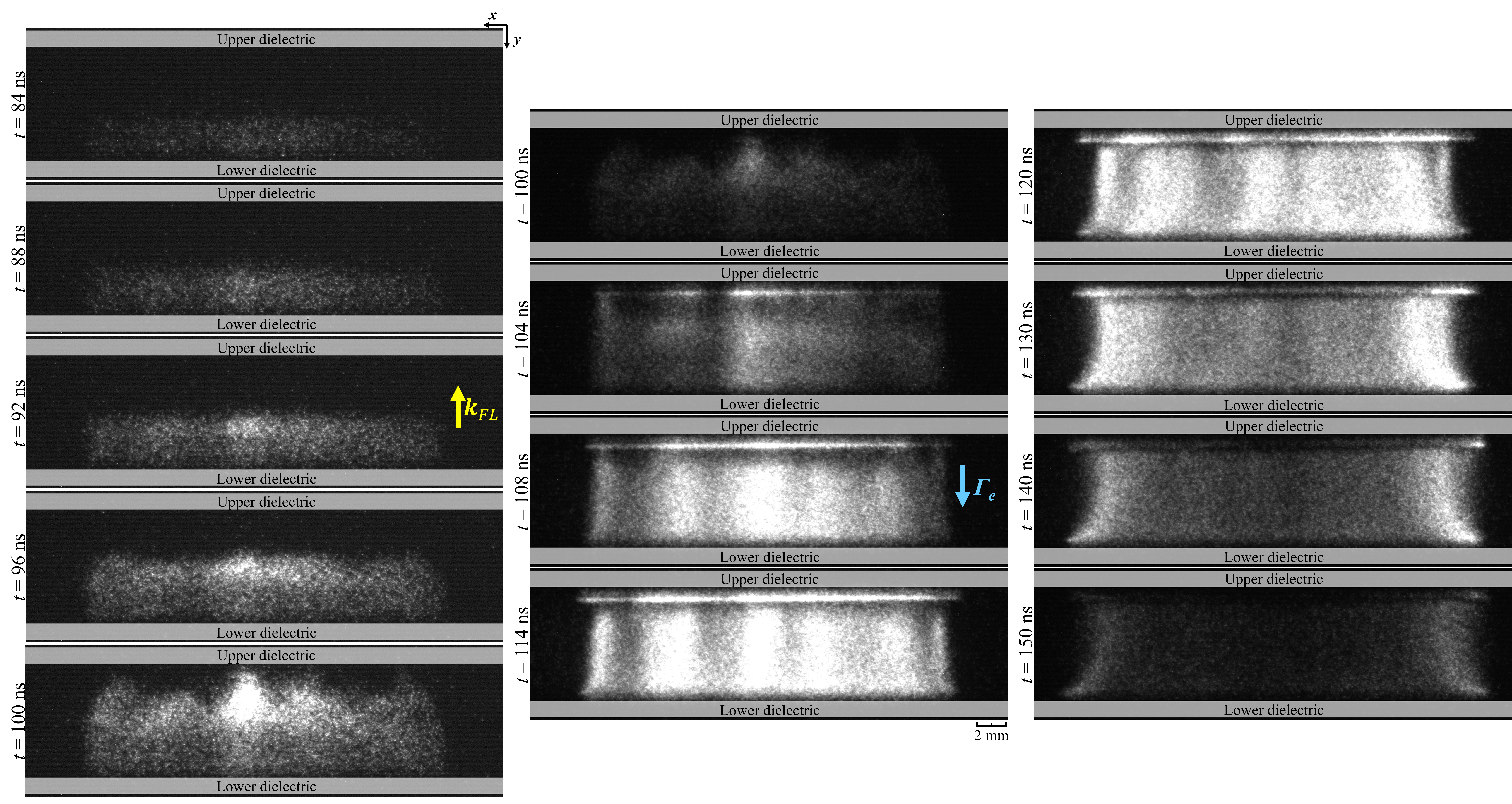}
	\caption{Selection of 1-ns iCCD images showing the development of the iDBD during NT ($t_0$ + 84, 88, 92, 96, 100, 104, 108, 114, 120, 130, 140, and 150 ns). Thick coloured arrows indicate the propagation direction of LPF and $\Gamma_e$ in yellow and blue, respectively. Measurements were conducted in pure \ce{NH3} at $10^4$ Pa, 6 kV$_{pp}$ and a gas gap \emph{g} of 7 mm.}
	\label{fig12}
\end{figure}

To support these attributions, a selection of images in figure \ref{fig12} presents the spatial evolution of the discharge-emitted light over time under the 7-mm gas gap condition. First, the luminous region near the anode slowly deforms towards the cathode, expanding by only about 2.5 mm along the \emph{y}-axis over 8 ns, as illustrated by the images from $t_0$ + 84 to 92 ns. During this increase, the LPF remains relatively flat along the \emph{x}-axis, exhibiting some similarities with a diffuse mode. These first three images appear to correspond well to slope II, as noted in figure \ref{fig11}a. Second, it is only from $t_0$ + 92 ns onward that the propagation accelerates, corresponding to slope III. This phase III is characterized by the formation of an intense and highly irregular LPF along both \emph{y}- and \emph{x}-axes, as highlighted with a sawtooth-like profile in the image at $t_0$ + 100 ns. This profile appears to be intrinsically linked to the columnar mode and is reminiscent of one of the results reported by C. Yao \emph{et al.} \cite{yao_characteristics_2018}. Indeed, and as shown with the following images, the discharge is characterized by columnar-like structures once the LPF have reached the cathode-like dielectric and is no longer defined. Moreover, under our conditions, the columns remain static along the \emph{x}-axis as illustrated by the images from $t_0$ + 108 to 120 ns. From $t_0$ + 120 ns approximately, the overall light intensity starts to dim indicating that the discharge is extinguishing concomitant of that of the current decreasing. Overall, this LPF profile is also linked to the self-organisation of charges along the \emph{x}-axis (and the \emph{z}-axis). This observational result supports the hypothesis of a competition between production and loss processes of charges, governed by the distribution of local electric field, which is itself influenced by the spatial charge distribution in the gas volume and on dielectrics during the whole period of the applied high voltage.

\section{Conclusion}

The temporal evolution of the iDBD-emitted light reveals two distinct phases. The first phase corresponds to the cathodic propagation of a front associated with ionisation and excitation, which is therefore luminous and referred to as LPF. This LPF forms near the anode and moves towards the cathode along the \emph{y}-axis, in the opposite direction of the global electron flux $\Gamma_e$. The second phase consists of an additional expansion along the \emph{x}-axis, parallel to the electrodes, occurring once the LPF reaches the cathode. This luminous expansion phase may reflect the establishment of an electron density gradient.

Moreover, the LPF temporal evolution is remarkably correlated with that of the current density, $J_{dis}$. The results demonstrate the presence of a proportionality factor of $1.5 \times$~$10^{-3}$ between the LPF average velocity and the $J_{dis}$ rise rate, regardless of the parameters in our study conditions (applied voltage \emph{V}, pressure \emph{p}, and gas gap \emph{g}). This correlation is observed only in the case of diffuse discharges. Further investigations are needed to determine the consistency of this proportionality factor value over other parameters. To this end, an experimental study of the nature of both the gas and the electrodes/dielectrics seems mandatory, completed further by an iDBD discharge modelling.

Furthermore, this correlation is questioned under columnar conditions, but it is very likely due to the overestimation of the interaction surfaces between the discharge and the dielectrics, which is difficult to determine in a columnar mode and leads to an underestimation of $J_{dis}$. Nonetheless, these results suggest that there is a relationship between the dynamics of LPF, associated with local excitation and de-excitation mechanisms in the volume, and $J_{dis}$, which integrates all the production and consumption of charge phenomena occurring within the volume at the surface level.

Nonetheless, in the case of diffuse discharges, the correlation between the LPF velocity and $J_{dis}$ rise rate supports the hypothesis of a critical charge density $n_{crit}$. The dynamics of the latter would govern the propagation of an excitation and ionisation front towards the cathode, which drives the $J_{dis}$ rise and allows one to link LPF propagation to electron flux $\Gamma_e$ – seemingly antagonistic since LPF moves against $\Gamma_e$, yet unified by $n_{crit}$.

%%%%%%%%%%%%%%%%%%%%%%%%%%%%%%%%%%%%%%%%%%%%%%%%%%%%%%%%%%%%%%%%%%%%%%%%%%%%%%%%%%%%%%%%%%
%Acknowledgments
%%%%%%%%%%%%%%%%%%%%%%%%%%%%%%%%%%%%%%%%%%%%%%%%%%%%%%%%%%%%%%%%%%%%%%%%%%%%%%%%%%%%%%%%%%
\phantomsection
\section*{Acknowledgments}
\addcontentsline{toc}{section}{Acknowledgments}

This work has been funded by the French Agence Nationale de la Recherche (ANR) under project SYNERGY (ANR-20-CE05-0013).

%%%%%%%%%%%%%%%%%%%%%%%%%%%%%%%%%%%%%%%%%%%%%%%%%%%%%%%%%%%%%%%%%%%%%%%%%%%%%%%%%%%%%%%%%%
%APPENDIX
%%%%%%%%%%%%%%%%%%%%%%%%%%%%%%%%%%%%%%%%%%%%%%%%%%%%%%%%%%%%%%%%%%%%%%%%%%%%%%%%%%%%%%%%%%

%%%%%%%%%%%%%%%%%%%%%%%%%%%%%%%%%%%%%%%%%%%%%%%%%%%%%%%%%%%%%%%%%%%%%%%%%%%%%%%%%%%%%%%%%%
%BIBLIOGRAPHY
%%%%%%%%%%%%%%%%%%%%%%%%%%%%%%%%%%%%%%%%%%%%%%%%%%%%%%%%%%%%%%%%%%%%%%%%%%%%%%%%%%%%%%%%%%
% le style aomalpha permet d'afficher un lien cliquable vers l'article cité %
%\cleardoublepage
\phantomsection

\bibliographystyle{aomalpha}
\bibliography{references.bib}
\addcontentsline{toc}{section}{References}

\end{document}